\documentclass[12pt]{article}

\usepackage[utf8]{inputenc}
\usepackage[margin=25truemm]{geometry}
\usepackage{physics}
\usepackage{amsthm}
\usepackage{mathtools}
\usepackage{ascmac}
\usepackage{latexsym}
\usepackage{indentfirst}
\usepackage{cancel}
\usepackage{comment}
\usepackage{graphicx}

\usepackage[colorlinks=true,linkcolor=blue,filecolor=blue,urlcolor=blue,citecolor=blue]{hyperref}

\usepackage{bigints}
\usepackage{amsmath,amssymb,bm}
\usepackage{braket}
\usepackage{esvect}

\newcommand {\beq}{\begin{equation}}
\newcommand {\eeq}{\end{equation}}
\newcommand {\beqa}{\begin{eqnarray}}
\newcommand {\eeqa}{\end{eqnarray}}

\newcommand{\UV}{\text{\tiny UV}}

\numberwithin{equation}{section}

\allowdisplaybreaks[1]

\begin{document}

\begin{titlepage}
\renewcommand{\thefootnote}{\fnsymbol{footnote}}
\begin{normalsize}
\begin{flushright}
\begin{tabular}{l}
\end{tabular}
\end{flushright}
\end{normalsize}

~~\\

\vspace*{0cm}
    \begin{Large}
       \begin{center}
         {Quantum error correction realized by the renormalization group  \\
         in scalar field  theories}
       \end{center}
    \end{Large}
\vspace{1cm}

\begin{center}
           Takaaki K{\sc uwahara}$^{1)}$\footnote
            {
e-mail address:
kuwahara@gauge.scphys.kyoto-u.ac.jp},
           Ryota N{\sc asu}$^{2)}$\footnote
            {
e-mail address:
nasu.ryota.17@shizuoka.ac.jp},
          Gota T{\sc anaka}$^{3)}$\footnote
            {
e-mail address:
gotanak@mi.meijigakuin.ac.jp}
           {and}
           Asato T{\sc suchiya}$^{2),4)}$\footnote
           {
e-mail address: tsuchiya.asato@shizuoka.ac.jp}
\\
      \vspace{1cm}

$^{1)}$
 {\it Department of Physics, Kyoto University}\\
                {\it Sakyo-ku, Kyoto 606-8502, Japan}\\
        \vspace{0.3cm}
$^{2)}$
 {\it Graduate School of Science and Technology, Shizuoka University}\\
               {\it 836 Ohya, Suruga-ku, Shizuoka 422-8529, Japan}\\
        \vspace{0.3cm}

$^{3)}$
{\it Institute for Mathematical Informatics, Meiji Gakuin University}\\
                {\it 1518 Kamikuratacho, Totsuka-ku, Yokohama 244-8539, Japan}
        \vspace{0.3cm}

 $^{4)}$
{\it Department of Physics, Shizuoka University}\\
                {\it 836 Ohya, Suruga-ku, Shizuoka 422-8529, Japan}\\

\end{center}

\vspace{2cm}

\begin{abstract}
\noindent
We demonstrate that quantum error correction is realized by the renormalization group
in scalar field theories. We construct $q$-level 
states by using coherent states
in the IR region. By acting on them the inverse of the unitary operator $U$ that describes the 
renormalization group flow of the ground state, we encode them into
states in the UV region. 
We find the situations in which the Knill-Laflamme condition is satisfied for 
operators that create coherent states.
We verify this to the first order in the perturbation theory.
This result suggests a general relationship between
the renormalization group and quantum error correction and 
should give insights into understanding the role played by them in the gauge/gravity correspondence.
\end{abstract}
\end{titlepage}
\vfil\eject

\setcounter{footnote}{0}

\section{Introduction}

Quantum error correction is important 
in the context of quantum information
such as quantum communication and quantum computing, 
while it has 
recently been recognized\cite{Almheiri:2014lwa}
that it plays a crucial role in the emergence of space-time
in the gauge/gravity correspondence\cite{Maldacena:1997re}. 
In this paper, we demonstrate that quantum error
correction is realized by the renormalization group in scalar field theories.

In quantum error correction, in order to protect quantum information possessed by 
span$(\{|i\rangle\})$ from errors, one encodes span$(\{|i\rangle\})$ into a larger Hilbert space
$\mathcal{H}$.
One defines the encoding map $W$ by
\begin{align}
|\tilde{i}\rangle = W |i\rangle \ ,
\label{encoding map}
\end{align}
where $|\tilde{i}\rangle$ are elements of $\mathcal{H}$ 
and $W^{\dagger}W=I$ is 
satisfied.
A subspace of $\mathcal{H}$, span$(\{|\tilde{i}\rangle\})$, 
is called the code subspace.

Suppose that the errors are described by error operators $\{E_a\}$ which are linear maps
in $\mathcal{H}$.
The condition for quantum error correction (the Knill-Laflamme condition)\cite{Knill:1996ny} is
given by
\begin{align}
\langle \tilde{i} | E_a^{\dagger}E_b |\tilde{j}\rangle = M_{ab}\delta_{ij} \ ,
\label{Knill-Laflamme condition 2}
\end{align}
where $M_{ab}$ are elements
of a Hermitian matrix and $\{|\tilde{i}\rangle\}$ are an 
orthonormal basis of the code subspace. 
The orthonormal basis can be expanded 
in terms of a non-orthonormal basis $\{\ket{\tilde{r}}\}$ as $\ket{\tilde{i}} = \sum_{r}\ket{\tilde{r}}T_{ri}$.
Then, it follows that
$\delta_{ij} = \braket{\tilde{i}|\tilde{j}} = \sum_{r,s}T^{*}_{ri}\braket{\tilde{r}|\tilde{s}}T_{sj}$. By acting $(T^{-1}_{ir})^*$ from left and $T^{-1}_{js}$ from right on \eqref{Knill-Laflamme condition 2}, one can rewrite
it as
\begin{align}
    \label{Knill-Laflamme condition 3}
    \langle \tilde{r} | E_a^{\dagger}E_b |\tilde{s}\rangle = M_{ab}\braket{\tilde{r}|\tilde{s}} \ .
\end{align}
In this paper, we use this expression for the quantum error correction condition.
The condition for quantum error condition is said to be satisfied in an approximate sense 
when (\ref{Knill-Laflamme condition 3}) holds up to a small quantity.

In \cite{Furuya:2020tzv, Furuya:2021lgx}, quantum error correction is realized by the renormalization group in so-called magic cMERA\cite{Zou:2019xbi}, which is 
a free scalar field theory whose action possesses a particular scale dependence, by using
coherent states.
In this paper, motivated by this work, we realize quantum error correction by the renormalization group
in scalar field theories including interactions.
We construct $q$-level states $|r\rangle$ $(r=0,1,\ldots,q-1)$, which form a non-orthonormal basis of $\text{span}(\{\ket{i}\})$, by using coherent states
in the IR region. 
We describe the renormalization group flow
of the ground state $|\Psi\rangle_{\Lambda}$ by a unitary operator $U$ as
\begin{align}
    \ket{\Psi}_{\Lambda}=U(\Lambda,\Lambda_{\UV})\ket{\Psi}_{\Lambda_{\UV}} \ , 
\end{align}
where $\Lambda$ is the effective cutoff and $\Lambda_{\UV}$
is the UV cutoff.
By acting on the $q$-level states
the inverse of the unitary operator $U$, we encode them into
states in the UV region. Namely, we identify $U^{\dagger}$ with $W$
in (\ref{encoding map}) and have
\begin{align}
|\tilde{r}\rangle = U^{\dagger}(\Lambda,\Lambda_{\UV}) | r\rangle \ .
\end{align}
We examine whether the condition \eqref{Knill-Laflamme condition 3}
is satisfied for a class of operators which correspond to $E_a$ in 
\eqref{Knill-Laflamme condition 2} 
to the first order in the perturbation theory.
We find the situations in which
the condition (\ref{Knill-Laflamme condition 2}) 
is satisfied in an approximate sense. 

This result suggests a general relationship 
between the renormalization group and
quantum error correction.
Furthermore, in the gauge/gravity correspondence,
the structure of quantum error correction
is seen 
in reconstructing operators in the bulk in terms of operators in the boundary theory
\cite{Almheiri:2014lwa}, and 
the bulk can be viewed as described by the IR region in the boundary theory
because the bulk direction
corresponds to the scale of the renormalization group. 
Thus, this result should give 
insights into understanding the role played by quantum error correction and the renormalization
group in the gauge/gravity correspondence.

This paper is organized as follows.
In section 2, we construct the unitary operator $U$ in scalar field theories and we examine the scaling of creation and annihilation operators
under the operation of $U$. In section 3, we construct $q$-level states in
the IR region of
the scalar field theories by using the coherent states and encode those states
into the UV region in terms of $U^{\dagger}$. Using the result in section 2, we
show in what situations the condition for quantum error condition (\ref{Knill-Laflamme condition 2})
is satisfied in an approximate sense.
Section 4 is devoted to the conclusion and discussion.

\section{Renormalization group flow of the ground state}
\subsection{The effective Hamiltonian for scalar field theory}
Throughout this paper, we consider scalar field theories in $d+1$ dimensions with UV cutoff and
use a shorthand notation:
\begin{align}
\int_p \equiv \int \frac{d^dp}{(2\pi)^d} \ , \;\;\;
\tilde{\delta}(p) = (2\pi)^d \delta^d(p)  \ ,
\end{align}
where $p$ stands for $d$-dimensional spatial momentum.
We denote the effective momentum cutoff by $\Lambda$.
By making a rescaling such as
\begin{align}
p \rightarrow \Lambda p \ , \;\;\; \varphi(p) \rightarrow \Lambda^{-\frac{d+1}{2}} \varphi(p) \ ,
\end{align}
we make all quantities dimensionless.

We define
a cutoff function $K(x)$ that possesses a profile in which it is almost one for $x<1$ and
rapidly damps for $x>1$ and introduce a momentum cutoff only
in the spatial directions by $K_p\equiv K(p^2)$.
Then, the effective Hamiltonian is given by
\begin{align}
	H_{\Lambda} &= H_{0,\Lambda} 
+ \alpha H_{\mathrm{\mathrm{int},\Lambda}}	\ , \label{fullH} 
\end{align}
where $H_{0,\Lambda}$ and $H_{\mathrm{int},\Lambda}$ are free and interaction
parts, respectively, and $\alpha$ is an expansion parameter.
$H_{0,\Lambda}$ and $H_{\mathrm{int},\Lambda}$ take the forms 
\begin{align}
	H_{0,\Lambda} &= \int_p K_p \left\{ \frac{1}{2} \pi(p)\pi(-p) + \frac{1}{2} K_p^{-2}\omega_{\Lambda,p}^2 \varphi(p) \varphi(-p) \right\} \ ,	\label{freeH}	\\
	H_{\mathrm{int},\Lambda} &= \frac{\delta m^2(\Lambda)}{2} \int_p \varphi(p)\varphi(-p)
    + \frac{\lambda(\Lambda)}{4!} \int_{p_i} \varphi(p_1)\varphi(p_2)\varphi(p_3)\varphi(p_4) \tilde{\delta} \left( \sum_i p_i \right) \ ,	\label{intH}
\end{align}
where 
$\omega_{\Lambda,p}=\sqrt{p^2+m^2\Lambda^{-2}}$. 
The flow equation for $\delta m^2(\Lambda)$  is given to the first order in 
$\lambda(\Lambda)$ by
\begin{align}
\label{scaling eq for delta m^2}
-\Lambda\frac{\partial}{\partial \Lambda}\delta m^2(\Lambda)=2\delta m^2(\Lambda)
-\frac{1}{2}\lambda(\Lambda) \int_p \left( \frac{d}{2\omega_{\Lambda,p}}
-\frac{p^2}{2\omega_{\Lambda,p}^3}\right) K_p  \ .
\end{align}

We denote the ground state of the Hamiltonian (\ref{fullH}) by $|\Psi\rangle_{\Lambda}$.
We describe the renormalization group flow of the ground state by a unitary operator $U(\Lambda,\Lambda_{\UV})$ as
\begin{align}
    \ket{\Psi}_{\Lambda}=U(\Lambda,\Lambda_{\UV})\ket{\Psi}_{\Lambda_{\UV}} \ , \label{definition of U}
\end{align}
where $\Lambda_{\UV}$ denotes UV cutoff,
and assume that $U(\Lambda,\Lambda_{\UV})$ can be represented as
\begin{align}
    U(\Lambda,\Lambda_{\UV})=\mathop{T}\exp\qty[\int_{\Lambda}^{\Lambda_{\UV}}\frac{d\Lambda'}{\Lambda'}X_{\Lambda'}] \ .\label{definition of disentangler}
\end{align}
Here $\mathop{T}$ is an ordering operator defined by
\begin{align}
T(X_{\Lambda}X_{\Lambda'})=\left\{
\begin{array}{ll}
X_{\Lambda}X_{\Lambda'} & \mbox{for} \; \Lambda < \Lambda' \\
X_{\Lambda'}X_{\Lambda} & \mbox{for} \; \Lambda > \Lambda'
\end{array}
\right.  \ .
\end{align}
As in the context of continuum MERA\cite{Haegeman:2011uy, Nozaki:2012zj}, it is natural to call an anti-Hermitian operator $X_{\Lambda}$ the disentangler because it removes entanglement and 
reduces degrees of freedom along with the renormalization group flow.

By acting $-\Lambda\partial_{\Lambda}$ on both sides of (\ref{definition of U}) and 
using (\ref{definition of disentangler}), we obtain
\begin{align}
    -\Lambda\partial_{\Lambda}\ket{\Psi}_{\Lambda}
=X_{\Lambda}\ket{\Psi}_{\Lambda}  \ .
\label{flow equation for ground state}
\end{align}
This is the flow equation for the ground state. 
If we obtain the scale dependence of the ground state in another way, 
we can calculate $X_{\Lambda}$ by using this equation.

We define the creation and annihilation operators at the scale $\Lambda$ by\footnote{
    It seems nontrivial whether there exist creation and annihilation operators that satisfy \eqref{commutation relation} and \eqref{definition of annihilation operator}. 
    In the following, we show this is indeed the case to the first order in the perturbation theory.    
}
\begin{align}
&[a_{\Lambda,p},a^{\dagger}_{\Lambda,p'}]=\tilde{\delta}(p-p') \ , \;\;
[a_{\Lambda,p},a_{\Lambda,p'}]=0 \ ,   \;\;
[a^{\dag}_{\Lambda,p},a^{\dag}_{\Lambda,p'}]=0 \ , 
\label{commutation relation} \\
&a_{\Lambda,p}|\Psi\rangle_{\Lambda}=0 \ .
\label{definition of annihilation operator}
\end{align}
Then, the renormalization group flow of the creation and annihilation operators is defined by
\begin{align}
& a_{\Lambda,p}=U(\Lambda,\Lambda_{\UV}) a_{\Lambda_{\UV},p} 
U(\Lambda,\Lambda_{\UV})^{\dagger} \ ,\nonumber\\
& a_{\Lambda,p}^{\dagger}=U(\Lambda,\Lambda_{\UV}) a_{\Lambda_{\UV},p}^{\dagger}
U(\Lambda,\Lambda_{\UV})^{\dagger}  \ ,
\label{transformation of creation and annihilation operators}
\end{align}
or, equivalently,
\begin{align}
&-\Lambda\partial_{\Lambda}a_{\Lambda,p}=[X_{\Lambda},a_{\Lambda,p}] \ , \nonumber\\
&-\Lambda\partial_{\Lambda}a_{\Lambda,p}^{\dagger}
=[X_{\Lambda},a_{\Lambda,p}^{\dagger}] \ .
\label{flow equations for creation and annihilation operators}
\end{align}
(\ref{commutation relation}) and (\ref{definition of annihilation operator}) are
preserved under (\ref{transformation of creation and annihilation operators}) 
and (\ref{flow equations for creation and annihilation operators}).
For later convenience, we also introduce linear combinations of the creation and annihilation operators as
\begin{align}
  &  a_{\Lambda,p}^{+} = a_{\Lambda,p}+a^{\dag}_{\Lambda,-p}{} \ , \\
  &  a_{\Lambda,p}^{-} = a_{\Lambda,p}-a^{\dag}_{\Lambda,-p}{} \ .
\end{align}
The flow equations for the above operators follow from 
(\ref{flow equations for creation and annihilation operators}):
\begin{align}
-\Lambda\partial_{\Lambda} a_{\Lambda}^{\pm}=\qty[X_{\Lambda},a_{\Lambda,p}^{\pm}]  \ .
\label{scaling_for_addition and subtraction op}
\end{align}

In this paper, we consider a perturbation theory in which
we expand $|\Psi\rangle_{\Lambda}$, $X_{\Lambda}$, $a_{\Lambda,p}$ and
$a_{\Lambda,p}^{\pm}$ in terms of $\alpha$
as follows:
\begin{align}
&|\Psi\rangle_{\Lambda}=|\Psi^{(0)}\rangle_{\Lambda}+\alpha |\Psi^{(1)}\rangle_{\Lambda}+
\alpha^2 |\Psi^{(2)}\rangle_{\Lambda}+ \cdots \ ,  \nonumber\\
&X_{\Lambda} = X^{(0)}_{\Lambda} +\alpha X^{(1)}_{\Lambda} + \alpha^2 X^{(2)}_{\Lambda} 
+\cdots \ , \nonumber\\
&a_{\Lambda,p}=a^{(0)}_{\Lambda,p}+\alpha a^{(1)}_{\Lambda,p}
+\alpha^2 a^{(2)}_{\Lambda,p}+\cdots \ , \nonumber\\
&a_{\Lambda,p}^{\pm}=a^{\pm(0)}_{\Lambda,p}+\alpha a^{\pm(1)}_{\Lambda,p}{}
+\alpha^2 a^{\pm(2)}_{\Lambda,p}{}+\cdots \ .
\label{perturbative expansion}
\end{align}
In the remaining part of this section, we use the $\varphi$-representation.

\subsection{Free field theory}
\label{free field theory}

In this subsection, we examine the free field theory, namely the zeroth order in $\alpha$.
The creation and annihilation operators for the free Hamiltonian 
$H_{0,\Lambda}$ are given by
\begin{align}
    a^{(0)}_{\Lambda,p}&=\frac{1}{\sqrt{2}}\qty(\sqrt{\frac{\omega_{\Lambda,p}}{K_{p}}}\varphi(p)+\sqrt{\frac{K_{p}}{\omega_{\Lambda,p}}}\frac{\delta}{\delta\varphi(-p)}) \ ,
\nonumber\\
    a^{(0)\dag}_{\Lambda,p}&=\frac{1}{\sqrt{2}}\qty(\sqrt{\frac{\omega_{\Lambda,p}}{K_{p}}}\varphi(-p)-\sqrt{\frac{K_{p}}{\omega_{\Lambda,p}}}\frac{\delta}{\delta\varphi(p)})\label{free creation and annihilation op} \ .
\end{align}
Then, the free Hamiltonian is rewritten as
\begin{align}
H_{0,\Lambda}= \int_p \omega_{\Lambda,p}
a^{(0)}_{\Lambda,p}{}^{\dagger}a^{(0)}_{\Lambda,p} + 
\frac{V}{2}\int_p\omega_{\Lambda,p}\ ,
\end{align}
where $V$ is the volume of space.
Then, the normalized wave functional for the ground state  
defined by $\Psi^{(0)}_{\Lambda}[\varphi]=\langle \varphi | \Psi^{(0)}\rangle_{\Lambda}$ 
satisfies $a^{(0)}_{\Lambda,p}\Psi^{(0)}_{\Lambda}=0$, from which we obtain
\begin{align}
    \Psi^{(0)}_{\Lambda}[\varphi]=\exp\qty[-\frac{1}{2}\int_{\vec{p}}\varphi(p)\frac{\omega_{\Lambda,p}}{K_{p}}\varphi(-p)+\frac{V}{4}\int_{p}\ln{\qty(\frac{2\omega_{\Lambda,p}}{K_p})}] \ . \label{WF for GS in FSFT2}
\end{align}

By acting $-\Lambda\partial_{\Lambda}$ on (\ref{WF for GS in FSFT2}), we obtain
\begin{align}
    -\Lambda\partial_{\Lambda}\Psi^{(0)}_{\Lambda}[\varphi]
 =\left(-\frac{1}{2}\int_{\vec{p}}\varphi(p)\frac{\dot{\omega}_{\Lambda,p}}{K_{p}}\varphi(-p)+\frac{V}{4}\int_{p}\frac{\dot{\omega}_{\Lambda,p}}{\omega_{\Lambda,p}}\right)
\Psi^{(0)}_{\Lambda}  \ ,
\label{calculating X 1}
\end{align}
where
\begin{align}
    \dot{\omega}_{\Lambda,p} = -\Lambda\partial_{\Lambda}\omega_{\Lambda,p} \ .
\end{align}
It follows from  (\ref{free creation and annihilation op}) that
\begin{align}
    \varphi(p)=\sqrt{\frac{K_{p}}{2\omega_{\Lambda,p}}}\qty(a^{(0)}_{\Lambda,p}+a^{(0)\dag}_{\Lambda,-p}) \ .
\end{align}
Substituting this into (\ref{calculating X 1}), we obtain
\begin{align}
    -\Lambda\partial_{\Lambda}\Psi^{(0)}_{\Lambda}
    =-\frac{1}{4}\int_{p}\frac{\dot{\omega}_{\Lambda,p}}{\omega_{\Lambda,p}}a^{(0)\dag}_{\Lambda,-p}a^{(0)\dag}_{\Lambda,p}\Psi^{(0)}_{\Lambda} \ .
\end{align}
Taking the anti-Hermiticity into account, we read off $X^{(0)}_{\Lambda}$ as
\begin{align}
    X^{(0)}_{\Lambda}=-\frac{1}{4}\int_{p}\frac{\dot{\omega}_{\Lambda,p}}{\omega_{\Lambda,p}}\qty(a_{\Lambda,-p}^{(0)\dag}a_{\Lambda,p}^{(0)\dag}
-a_{\Lambda,p}^{(0)}a^{(0)}_{\Lambda,-p}) \ , 
\label{X^(0)}
\end{align}
where we use $a^{(0)}_{\Lambda,p}\Psi^{(0)}_{\Lambda}=0$.

Finally, we derive the scaling of the creation and annihilation operators
in the free field theory.
While we can easily read it off from (\ref{free creation and annihilation op}),
we derive it by using the disentangler as a preparation for the analysis of the interacting theory.
By using (\ref{X^(0)}), we calculate the zeroth order 
of (\ref{scaling_for_addition and subtraction op}) in $\alpha$ as
\begin{align}
-\Lambda\partial_{\Lambda}a_{\Lambda,p}^{\pm(0)}
  & =  \qty[X_{\Lambda}^{(0)},a_{\Lambda,p}^{\pm(0)}]  \nonumber\\
 &= \pm\frac{1}{2} 
\frac{\dot{\omega}_{\Lambda,p}}{\omega_{\Lambda,p}}
a_{\Lambda,p}^{\pm(0)} \ ,
\end{align}
from which, we obtain the scaling for $a^{\pm(0)}$ as
\begin{align}
    a_{\Lambda,p}^{+(0)}&=\sqrt{\frac{\omega_{\Lambda,p}}{\omega_{\UV,p}}}a_{\UV,p}^{+(0)} 
\ , \\
    a_{\Lambda,p}^{-(0)}&=\sqrt{\frac{\omega_{\UV,p}}{\omega_{\Lambda,p}}}a_{\UV,p}^{-(0)} 
\ ,
\end{align}
where $\omega_{\UV,p}$ and $a^{\pm(0)}_{\UV,p}$ stand for
$\omega_{\Lambda,p}$ and $a^{\pm(0)}_{\Lambda,p}$ with $\Lambda=\Lambda_{\UV}$,
respectively.

\subsection{The first order in the perturbation theory}
\label{subsec: the 1st order in perturbation theory}
In this subsection, we examine the first order in the perturbation theory.
In appendix A, we calculate $\Psi_{\Lambda}^{(1)}$.
The result is\footnote{$\Psi^{(0)}_{\Lambda}+\alpha \Psi^{(1)}_{\Lambda}$ is a perturbative solution to the exact renormalization group equation for the wave functional of the ground state in scalar field theories derived in \cite{Kuwahara:2022nlm} .}
\begin{align}
\Psi^{(1)}_{\Lambda} = A_{\Lambda} \Psi^{(0)}_{\Lambda} 
\label{Psi^(1)}
\end{align}
with
\begin{align}
A_{\Lambda}
= &- \frac{\lambda}{4!} \int_{k_1, \dots , k_4} \frac{ \tilde{\delta}(k_1 + \cdots k_4) }{\omega_{\Lambda,k_1} + \cdots + \omega_{\Lambda,k_4}}
			\prod_{i=1}^4 \sqrt{\frac{K_{k_i}}{2\omega_{\Lambda,k_i}}} a^{(0)\dagger}_{\Lambda,k_i}\notag\\
 &\qquad\qquad
 - \left( \frac{\delta m^2}{2} + \frac{\lambda}{4!} \int_p \frac{6K_p}{2\omega_{\Lambda,p}} \right)
\int_k \frac{1}{2\omega_{\Lambda,k}}  \frac{K_k}{2\omega_{\Lambda,k}} a^{(0)\dagger}_{\Lambda,k} a^{(0)\dagger}_{\Lambda,-k}	  \nonumber\\
&\qquad\qquad - (\mbox{Hermitian conjugate}) \ .
\label{A}
\end{align}
Note that $A_\Lambda$ is an anti-Hermitian operator.
By expanding (\ref{flow equation for ground state}), we obtain 
\begin{align}
    -\Lambda\partial_\Lambda \Psi^{(0)}_{\Lambda}&=X^{(0)}_{\Lambda}
\Psi^{(0)}_\Lambda \ ,  \label{scaling eq for Psi^(0)} \\
    -\Lambda\partial_\Lambda\Psi^{(1)}_{\Lambda}&=X^{(0)}_{\Lambda}
\Psi^{(1)}_\Lambda+X^{(1)}_{\Lambda}\Psi^{(0)}_\Lambda \ .
\label{scaling eq for Psi^(1)}
\end{align}
Substituting (\ref{Psi^(1)}) and (\ref{scaling eq for Psi^(0)}) into
(\ref{scaling eq for Psi^(1)}) leads to
\begin{align}
    -\Lambda\partial_{\Lambda}A_{\Lambda}
=X^{(1)}_{\Lambda}+[X^{(0)}_{\Lambda},A_{\Lambda}] \ .
\label{flow equation for A}
\end{align}
One can determine $X^{(1)}_{\Lambda}$ if $A_{\Lambda}$ is known.
In appendix B, we give the explicit form of $X^{(1)}_{\Lambda}$.

From 
(\ref{flow equations for creation and annihilation operators}), we obtain
\begin{align}
&   -\Lambda\partial_\Lambda a^{(0)}_{\Lambda,p}
=\qty[X^{(0)}_{\Lambda},a^{(0)}_{\Lambda,p}]  \label{scaling for a^(0)} \ , \\
 &   -\Lambda\partial_ {\Lambda}a^{(1)}_{\Lambda,p}=\qty[X^{(1)}_{\Lambda},a^{(0)}_{\Lambda,p}]+\qty[X^{(0)}_{\Lambda},a^{(1)}_{\Lambda,p}] \ .
\label{scaling for a1}
\end{align}
In the following, we show that
\begin{align}
    a^{(1)}_{\Lambda,p}=[A_{\Lambda},a^{(0)}_{\Lambda,p}] \ .
\label{solution of a1 scaling}
\end{align}
First, we verify that
(\ref{solution of a1 scaling}) satisfies (\ref{commutation relation}) and 
(\ref{definition of annihilation operator}) up to the first order in $\alpha$.
We calculate $[a_{\Lambda,p},a^{\dag}_{\Lambda,p'}]$ as follows:
\begin{align}
[a_{\Lambda,p},a^{\dag}_{\Lambda,p'}]
=[a_{\Lambda,p}^{(0)},a^{(0)\dag}_{\Lambda,p'}]
+\alpha [a_{\Lambda,p}^{(0)},[A_{\Lambda},a^{(0)\dag}_{\Lambda,p'}]]
+\alpha[[A_{\Lambda},a^{(0)}_{\Lambda,p}],a^{(0)\dag}_{\Lambda,p'}]
+\mathcal{O}(\alpha^2) \ ,
\label{calculation of commutation relation}
\end{align}
where we used the anti-Hermiticity of $A_{\Lambda}$.
The first term in the RHS of (\ref{calculation of commutation relation})
gives $\tilde{\delta}(p-p')$, while the second and third terms 
reduce to $\alpha[A_{\Lambda},[a_{\Lambda,p}^{(0)},a^{(0)\dag}_{\Lambda,p'}]]=0$ due
to the Jacobi identity. In a similar manner, we can show that $[a_{\Lambda,p},a_{\Lambda,p'}]= [a_{\Lambda,p}^{\dag},a_{\Lambda,p'}^{\dag}] =0$.
We calculate $a_{\Lambda,p} \Psi_{\Lambda}$ as
\begin{align}
a_{\Lambda,p}\Psi_{\Lambda}
=a^{(0)}_{\Lambda,p}\Psi^{(0)}_{\Lambda}
+\alpha a_{\Lambda,p}^{(0)}A_{\Lambda}\Psi^{(0)}_{\Lambda}
+\alpha [A_{\Lambda},a^{(0)}_{\Lambda,p}]\Psi^{(0)}_{\Lambda} 
+\mathcal{O}(\alpha^2)  \ .
\end{align}
The first term in RHS vanishes as shown in section \ref{free field theory},
while the second and third terms cancel each other.
Next, acting $-\Lambda\partial_\Lambda$ on both sides of (\ref{solution of a1 scaling}), 
we verify that (\ref{solution of a1 scaling}) is a solution to (\ref{scaling for a1}): 
\begin{align}
    -\Lambda\partial_{\Lambda}a^{(1)}_{\Lambda,p}
    &=[-\Lambda\partial_{\Lambda}A_{\Lambda},a^{(0)}_{\Lambda,p}] 
+[A_{\Lambda},-\Lambda\partial_{\Lambda}a^{(0)}_{\Lambda,p}]
 \nonumber\\
    &=\qty[X^{(1)}_{\Lambda}+[X^{(0)}_{\Lambda},A_{\Lambda}],a^{(0)}_{\Lambda,p}]
+\qty[A_{\Lambda},[X^{(0)}_{\Lambda},a^{(0)}_{\Lambda,p}]]
\nonumber\\
    &=\qty[X^{(1)}_{\Lambda},a^{(0)}_{\Lambda,p}]+\qty[X^{(0)}_{\Lambda},a^{(1)}_{\Lambda,p}]
\ ,
\end{align}
where we used the Jacobi identity in the last equality.
Thus, we showed (\ref{solution of a1 scaling}).

\section{Quantum error correction by the renormalization group}
\subsection{Encoding $q$-level states}
In this subsection, we do \textit{not} restrict ourselves to the free field theory.
We use only the properties of the creation and annihilation operators
(\ref{commutation relation}) and (\ref{definition of annihilation operator}).
In order to realize
a $q$-level system in scalar field theories, following \cite{Furuya:2021lgx}, we use coherent states defined by
\begin{align}
    \ket{f}_{\Lambda}=\exp\qty[\int_{p}\qty(f(p)a_{\Lambda,-p}^{\dag}-f^{*}(-p)a_{\Lambda,p})]\ket{\Psi}_{\Lambda} \ ,
\end{align}
where $f$ is an arbitrary function. Note that 
\begin{align}
a_{\Lambda,p}|f\rangle_{\Lambda} = f(p)|f\rangle_{\Lambda} \ .
\label{eigenstate of annihilation operator}
\end{align}
The inner product between coherent states is given by
\begin{align}
    \sideset{_{\Lambda}}{_{\Lambda}}{\mathop{\Braket{f'|f}}}=\exp\qty[-\frac{1}{2}\int_p\qty(\abs{f'(p)}^2-2f'^{*}(p)f(p)+\abs{f(p)}^2)]\ ,\label{inner product of coherent state}
\end{align}
which implies that  
${}_{\Lambda} \langle f | f \rangle_{\Lambda}=1$.
We construct $q$-level states by choosing $f=rf_0$ with $f_0$ being a real function
and $r=0,1,\ldots,q-1$ : 
\begin{align}
        \Ket{rf_{0}}_{\Lambda}
        &=\exp\qty[-r\int_{p}f_{0}(-p)\qty(a_{\Lambda,p}-a^{\dag}_{\Lambda,-p})]
           \ket{\Psi}_{\Lambda}  \nonumber\\
       & =\exp \qty[-r\int_{p} f_{0}(-p) a_{\Lambda,p}^- ]  \ket{\Psi}_{\Lambda}  \ .
\end{align}
The inner product between these states is given by
\begin{align}
{}_{\Lambda}\langle r'f_0 | r f_0 \rangle_{\Lambda}
        =\exp\qty[-\frac{1}{2}(r-r')^2\int_{p}\abs{f_{0}(p)}^2] \ .
\label{orthonormality}
\end{align}
Note that these states form an orthonormal basis in an approximate way when $\int_p\abs{f_{0}}^2$ is large enough.
When $f_0(x)$ is localized around $x=x_0$, the $q$-level states are realized locally.

We encode the $q$-level states into states in the UV region as
\begin{align}
    \Ket{rf_{0}}_{\UV}=U^{\dag}(\Lambda,\Lambda_{\UV})\Ket{rf_{0}}_{\Lambda},
\end{align}
where $U^{\dag}(\Lambda,\Lambda_{\UV})$ is 
the inverse of the unitary operator defined in (\ref{definition of U}).
This equation implies that we encode information defined in the IR region
with small $\Lambda$ into 
the UV region in terms of the inverse of the renormalization group\footnote{
Note that $U(\Lambda,\Lambda_{\UV})$ does \textit{not} necessarily give the renormalization group flow of $|rf_0\rangle_{\Lambda}$, 
since it is defined as giving that of the ground state. 
We just define the encoding of $|rf_0\rangle_{\Lambda}$
by $U^{\dag}(\Lambda,\Lambda_{\UV})$.
}.

In what follows, we consider error operators $D[g]$ defined in the IR region
\begin{align}
    D[g]=\exp\qty[\int_{p}g(-p)\qty(a_{\Lambda,p}-a^{\dag}_{\Lambda,-p})]
=\exp\qty[\int_{p}g(-p)a_{\Lambda,p}^-] \ ,
\end{align}
where $g$ is an arbitrary real function. 
Note that $D[g]$ is an operator that generates a coherent state in the IR region; namely, $D[g]|\Psi\rangle_{\Lambda}$ is a coherent state.
We examine whether the quantum error correction condition or the Knill-Laflamme condition\cite{Knill:1996ny}
\begin{align}
    \sideset{_{\UV}}{}{\mathop{\Bra{r'f_{0}}}}D^{\dag}[g]D[h]\Ket{rf_{0}}_{\UV} = M[g,h]\sideset{_{\UV}}{_{\UV}}{\mathop{\Braket{r'f_0|rf_0}}}\ ,\label{KL condition as exp value}
\end{align}
with $M[g,h]$ being a Hermitian matrix on the functional vector space 
is approximately satisfied.
Namely, we see that
span$(\{\Ket{rf_{0}}_{\UV}\})$ gives a code subspace that is correctable for the errors caused by $D[g]$.
In order to show this, it is enough to calculate
\begin{align}
    \sideset{_{\UV}}{}{\mathop{\Bra{r'f_{0}}}}D[g]\Ket{rf_{0}}_{\UV},\label{error exp value}
\end{align}
for any real functions $g$, because $D^{\dag}[g]D[h]=D[h-g]$.

\subsection{Free field theory}
In this subsection, we examine whether 
the error correction condition is satisfied
 in the free scalar field theory.  (\ref{error exp value}) is calculated as follows.
\begin{align}
\label{eq: matrix elements of D[g]}
    \sideset{_{\UV}}{}{\mathop{\Bra{r'f_{0}}}}&D[g]\Ket{rf_{0}}_{\UV}\nonumber\\ 
    &=\sideset{_{\UV}}{}{\mathop{\Bra{r'f_{0}}}}\exp\qty[\int_{p}g(-p)a^{-(0)}_{\Lambda,p}]\Ket{rf_{0}}_{\UV} \nonumber\\
    &=\sideset{_{\UV}}{}{\mathop{\Bra{r'f_{0}}}}\exp\qty[\int_{p}g(-p)\sqrt{\frac{\omega_{\UV,p}}{\omega_{\Lambda,p}}}a^{-(0)}_{\UV,p}]\Ket{rf_{0}}_{\UV}\nonumber\\
    &=\sideset{_{\UV}}{}{\mathop{\Bra{r'f_{0}}}}\exp\qty[-\int_{p}\qty(rf_{0}(-p)-g(-p)\sqrt{\frac{\omega_{\UV,p}}{\omega_{\Lambda,p}}})a^{-(0)}_{\UV,p}]\Ket{\Psi}_{\UV}\nonumber\\
    &=\sideset{_{\UV}}{_{\UV}}{\mathop{\Braket{r'f_{0}|rf_{0}-\sqrt{\frac{\omega_{\UV}}{\omega_{\Lambda}}}g}}}\nonumber\\
    &=\exp[-\frac{1}{2}\int_{p}\qty(-2(r-r')\sqrt{\frac{\omega_{\UV,p}}{\omega_{\Lambda,p}}}g(-p)f_{0}(p)+\frac{\omega_{\UV,p}}{\omega_{\Lambda,p}}\abs{g(p)}^2)]\sideset{_{\UV}}{_{\UV}}{\mathop{\Braket{r'f_0|rf_0}}}
\ ,
\end{align}
where we used (\ref{inner product of coherent state}).
Note that $\omega_{\Lambda,p}=\sqrt{p^2+(m/\Lambda)^2}$.

In the last line in 
\eqref{eq: matrix elements of D[g]}, 
the first term in the exponent, which is proportional to $r-r'$, can prevent the Knill-Laflamme condition from being satisfied. 
Thus, it is convenient to rewrite 
\eqref{eq: matrix elements of D[g]} as
\begin{align}
       \sideset{_{\UV}}{}{\mathop{\Bra{r'f_{0}}}}&D[g]\Ket{rf_{0}}_{\UV} = \exp\qty[-\frac{1}{2}\int_p\frac{\omega_{\UV,p}}{\omega_{\Lambda,p}}\abs{g(p)}^2] \sideset{_{\UV}}{_{\UV}}{\mathop{\Braket{r'f_0|rf_0}}} + \varepsilon_{\Lambda}
\end{align}
with
\begin{align}
    \label{epsilon_free}
    \varepsilon_{\Lambda} 
    &=
    \exp\qty[-\frac{1}{2}\int_p\frac{\omega_{\UV,p}}{\omega_{\Lambda,p}}\abs{g(p)}^2] 
    \sideset{_{\UV}}{_{\UV}}{\mathop{\Braket{r'f_0|rf_0}}}
    \qty[\exp[\int_{p}\qty((r-r')\sqrt{\frac{\omega_{\UV,p}}{\omega_{\Lambda,p}}}g(-p)f_{0}(p))]-1]
    \ .
\end{align}
When $\varepsilon_{\Lambda}$ is small,
the Knill-Laflamme condition is satisfied 
in an approximate sense.
Note that $\varepsilon_{\Lambda}=0$ for $r=r'$.
There are three cases where $\varepsilon_{\Lambda}$
is small for $r\neq r'$: (i)
When $\int_p |f_0(p)|^2$ is large enough,  
$\{|rf_0\rangle_{\UV}\}$ almost forms an orthonormal basis
so that $\sideset{_{\UV}}{_{\UV}}{\mathop{\Braket{r'f_0|rf_0}}} $ 
is small.
(ii) The overlap between $f_0$ and $g$ is small.
(iii) $m/\Lambda$ is small.
In case (ii) or (iii), $\varepsilon_{\Lambda}$ is given by
\begin{align}
    \begin{aligned}
    \varepsilon_{\Lambda} 
    &=
    (r-r')\int_q g(-q)\sqrt{\frac{\omega_{\UV,q}}{\omega_{\Lambda,q}}}f_{0}(q) \sideset{_{\UV}}{_{\UV}}{\mathop{\Braket{r'f_0|rf_0}}}\exp\qty[-\frac{1}{2}\int_p\frac{\omega_{\UV,p}}{\omega_{\Lambda,p}}\abs{g(p)}^2] \ .
    \end{aligned}
\end{align}
In particular, in case (iii), we can expand $\varepsilon_{\Lambda}$ in terms of $\Lambda/m$ as
\begin{align}
\label{epsilon_lambda_expanded}
        \begin{aligned}
    \varepsilon_{\Lambda} 
    &=
    (r-r')\sqrt{\frac{\Lambda}{m}}\int_q g(-q)\omega_{\UV,q}^{1/2}f_{0}(q)\sideset{_{\UV}}{_{\UV}}{\mathop{\Braket{r'f_0|rf_0}}} + \mathcal{O}((\Lambda/m)^{3/2})) \ .\\
    \end{aligned}  
\end{align}
We see from \eqref{epsilon_lambda_expanded} that the $\varepsilon_{\Lambda}$ scales as $(\Lambda/m)^{1/2}$ and that 
the combination of the orthogonality of $\{|rf_0\rangle_{\UV}\}$, the locality of the error (the  overlap between $f_0$ and $g$) and the energy scale $\Lambda/m$ determines how accurately the Knill-Laflamme condition is satisfied.


\subsection{Perturbation theory}
In this subsection, 
we examine whether the Knill-Laflamme condition is satisfied to the first order in the perturbation theory.
In order to calculate  (\ref{error exp value}), we begin with 
representing the error operator up to the first order in $\alpha$
in terms of $a^{\pm}_{\UV,p}$ by using
(\ref{solution of a1 scaling}) and (\ref{A}). 
The details of the calculation are presented
in appendix C.
The result is
\begin{align}
    D[g]=\exp \qty[ \int_{p} g(-p) a^{-}_{\Lambda,p}] = \exp \qty[X+\alpha Y]  \label{perturbative_error_op}
\end{align}
with
\begin{align}
    X&= \int_p g(-p)\sqrt{\frac{\omega_{\UV,p}}{\omega_{\Lambda,p}}}a^{-}_{\UV,p} \ , \label{Xop}\\
    Y
    & =\int_p g(-p)\bigg\{- \frac{\lambda}{4!} \int _{k_1 k_2 k_3} C_{1}(k_1, k_2, k_3; p) a^{-}_{\UV,-k_1}a^{-}_{\UV,-k_2}a^{-}_{\UV,-k_3}  \nonumber\\
        &- \frac{\lambda}{8} \int_{k_1 k_2 k_3} C_{2}(k_1, k_2; k_3; p)a^{+}_{\UV,-k_1}a^{+}_{\UV,-k_2}a^{-}_{\UV,-k_3}  \nonumber\\
        &- \frac{\lambda}{4} \int_{k} C_{3}(k; p)a^+_{\UV,p} - C_{4}(p)a^{-}_{\UV,p}\bigg\}  \ , \label{Yop}
\end{align}
where
\begin{align}
    &\begin{aligned}
        &C_{1}(k_1,k_2,k_3;p)\\
        &\quad=\tilde{\delta}(k_{1} + \dots + p)\left(\frac{1}{\omega_{\Lambda,1}+\cdots+\omega_{\Lambda,p}}\qty(\prod_{i=1}^{3} \sqrt{\frac{K_i}{2\omega_{\Lambda,i}}}) \sqrt{\frac{K_p}{2\omega_{\Lambda,p}}}\sqrt{\frac{\omega_{\UV,1}\omega_{\UV,2}\omega_{\UV,3}}{\omega_{\Lambda,1}\omega_{\Lambda,2}\omega_{\Lambda,3}}}\right.\\
        &\qquad\qquad\quad\hphantom{\tilde{\delta}(k_{1} + \dots + p)}
        \left.-\frac{1}{\omega_{\UV,1}+\cdots+\omega_{\UV,p}}\qty(\prod_{i=1}^{3} \sqrt{\frac{K_i}{2\omega_{\UV,i}}})\sqrt{\frac{K_p}{2\omega_{\UV,p}}}\sqrt{\frac{\omega_{\UV,p}}{\omega_{\Lambda,p}}}\right),
    \end{aligned}\label{C_1}\\
    &\begin{aligned}
        &C_{2}(k_1,k_2;k_3;p)\\
        &\quad=\tilde{\delta}(k_1 + \dots + p)\left(\frac{1}{\omega_{\Lambda,1}+\cdots+\omega_{\Lambda,p}} \qty( \prod_{i=1}^{3} \sqrt{\frac{K_i}{2\omega_{\Lambda,i}}}) \sqrt{\frac{K_p}{2\omega_{\Lambda,p}}}\sqrt{\frac{\omega_{\Lambda,1}\omega_{\Lambda,2}}{\omega_{\UV,1}\omega_{\UV,2}}} \sqrt{\frac{\omega_{\UV,3}}{\omega_{\Lambda,3}}}\right.\\
        &\qquad\qquad\quad\hphantom{\tilde{\delta}(k_{1} + \dots + p)}\left.
        -\frac{1}{\omega_{\UV,1}+\cdots+\omega_{\UV,p}}\qty(\prod_{i=1}^{3} \sqrt{\frac{K_i}{2\omega_{\UV,i}}})\sqrt{\frac{K_p}{2\omega_{\UV,p}}}\sqrt{\frac{\omega_{\UV,p}}{\omega_{\Lambda,p}}}\right),
    \end{aligned}\label{C_2}\\
    &\begin{aligned}
        C_{3}(k;p)=\frac{1}{2\omega_{\Lambda,p} + 2\omega_{\Lambda,k}} \frac{K_p}{2\omega_{\Lambda,p}} \frac{K_k}{2\omega_{\Lambda,k}}\sqrt{\frac{\omega_{\Lambda,p}}{\omega_{\UV,p}}}
        -\frac{1}{2\omega_{\UV,p} + 2\omega_{\UV,k}} \frac{K_p}{2\omega_{\UV,p}} \frac{K_k}{2\omega_{\UV,k}}\sqrt{\frac{\omega_{\UV,p}}{\omega_{\Lambda,p}}},
    \end{aligned}\label{C_3}\\
    &\begin{aligned}
        C_{4}(p)=\sqrt{\frac{\omega_{\UV,p}}{\omega_{\Lambda,p}}}\qty[
        \qty(\frac{\delta m^2_{\Lambda}}{2} + \frac{\lambda}{4!} \int_{q} \frac{6 K_{q}}{2\omega_{\Lambda,q}}) \frac{K_p}{2\omega^2_{\Lambda,p}}
        - \qty( \frac{\delta m^2_{\UV}}{2} + \frac{\lambda}{4!} \int_{q}\frac{6 K_{q}}{2\omega_{\UV,q}})\frac{K_p}{2\omega^{2}_{\UV,p}}] \ .
    \end{aligned}\label{C_4}
\end{align}
Here $\omega_{\Lambda,i}$, $\omega_{\UV,i}$ and $K_i$ 
stand for $\omega_{\Lambda,k_i}$, $\omega_{\UV,k_i}$ and $K_{k_i}$, respectively.
Note that 
$C_{1}(k_1,k_2,k_3;p)$ is symmetric with respect to the permutation of $k_1,\,k_2$ and $k_3$ 
while $C_{2}(k_1,k_2;k_3;p)$ is symmetric with respect to the permutation of $k_1$ and $k_2$.

By using the Baker-Campbell-Hausdorff formula, up to the first order of perturbation, we obtain
\begin{align}
    D[g]
=(1+\alpha Z)e^{X} \ ,
\label{Z}
\end{align}
where
\begin{align}
    Z=Y+\frac{1}{2}\qty[X,Y]-\frac{1}{12}\qty[X,\qty[X,Y]] \ .
\end{align}
We can calculate each commutator in (\ref{Z}) as follows:
\begin{align}
        \qty[X,Y]
       = &-\frac{\lambda}{2}\int_{p,\ell,k_2 k_3}\sqrt{\frac{\omega_{\UV,p}}{\omega_{\Lambda,p}}}g(-p)g(-\ell)C_{2}(p, k_2; k_3; \ell)a^+_{\UV,-k_2}a^+_{\UV,-k_3} \nonumber\\
        &\quad- \frac{\lambda}{2} \int_{p,k}\sqrt{\frac{\omega_{\UV,p}}{\omega_{\Lambda,p}}}g(-p)g(p) C_{3}(k; -p) \ , \\
     \qty[X,\qty[X,Y]]
= & -\lambda\int_{p,q,\ell,k}\sqrt{\frac{\omega_{\UV,q}\omega_{\UV,p}}{\omega_{\Lambda,q}\omega_{\Lambda,p}}}g(-p)g(-\ell)g(-q)C_{2}(p, q; k; \ell)a^+_{\UV,-k} \ .
\end{align}
Then, up to the first order in $\alpha$, (\ref{error exp value}) reduces to
\begin{align}
    &\sideset{_{\UV}}{}{\mathop{\Bra{r'f_{0}}}} D[g] \Ket{rf_{0}}_{\UV}\notag\\ 
    &=\sideset{_{\UV}}{_{\UV}}{\mathop{\Braket{r'f_{0}|rf_{0}-\sqrt{\frac{\omega_{\UV}}{\omega_{\Lambda}}}g}}} + \alpha \sideset{_{\UV}}{_{\UV}}{\mathop{\mel**{r'f_{0}}{\left( Y + \frac{1}{2} [X,Y] - \frac{1}{12} [X,[X,Y]] \right)}{rf_{0}-\sqrt{\frac{\omega_{\UV}}{\omega_{\Lambda}}}g}}} \ ,
\label{error exp value 2}
\end{align}
where
\begin{align}
    \Ket{rf_{0}-\sqrt{\frac{\omega_{\UV}}{\omega_{\Lambda}}}g}_{\UV}= e^X \Ket{rf_{0}}_{\UV} \ .
\end{align}
The calculation of (\ref{error exp value 2}) is given in appendix C.

The final result is 
\begin{align}
    \sideset{_{\UV}}{}{\mathop{\Bra{r'f_{0}}}}&D[g]\Ket{rf_{0}}_{\UV} = \exp\qty[-\frac{1}{2}\int_p\frac{\omega_{\UV,p}}{\omega_{\Lambda,p}}\abs{g(p)}^2](1+\alpha \Gamma) \sideset{_{\UV}}{_{\UV}}{\mathop{\Braket{r'f_0|rf_0}}} + \varepsilon_{\text{free},\Lambda}+\alpha\varepsilon_{\text{int},\Lambda} \ .
\end{align}
Here
$\varepsilon_{\text{free},\Lambda}$ is given in \eqref{epsilon_free}, and $\varepsilon_{\text{int},\Lambda}$ is given by
\begin{align}
    \varepsilon_{\text{int},\Lambda} = \qty(\exp \qty[-\frac{1}{2}\int_p \frac{\omega_{\UV,p}}{\omega_{\Lambda,p}} \abs{g(p)}^2]\sideset{_{\UV}}{_{\UV}}{\mathop{\Braket{r'f_0|rf_0}}} + \varepsilon_{\text{free},\Lambda})\qty(\Delta + \Gamma) \ .
\end{align}
$\Gamma$ and $\Delta$ are defined by
\begin{align}
    \Gamma =& \frac{\lambda}{4!} \int_{p,k_1,k_2,k_3} g(-p) C_{1} (k_1, k_2, k_3 ; p)  \sqrt{\frac{\omega_{\UV,1}\omega_{\UV,3}\omega_{\UV,2}}{\omega_{\Lambda,1}\omega_{\Lambda,2}\omega_{\Lambda,3}}} g(-k_1)g(-k_2) g(-k_3)\nonumber\\
    &- \frac{\lambda}{8} \int_{p,k_1,k_2} g(-p) C_{1} (k_1, k_2, -k_1 -k_2 ; p) \sqrt{\frac{\omega_{\UV,2}}{\omega_{\Lambda,2}}} g(-k_2)  \nonumber\\
    &+ \frac{\lambda}{8} \int_{p,k_1,k_2,k_3} g(-p) C_2(k_1, k_2; k_3; p)\sqrt{\frac{\omega_{\UV,1} \omega_{\UV,2} \omega_{\UV,3}}{\omega_{\Lambda,1} \omega_{\Lambda,2} \omega_{\Lambda,3}}} g(-k_1) g(-k_2) g(-k_3) \nonumber\\
    &- \frac{\lambda}{4} \int_{p,k_1,k_2,k_3} g(-p) C_2(k_1, k_2; -k_1; p) \sqrt{\frac{\omega_{\UV,2}}{\omega_{\Lambda,2}}} g(-k_2) \nonumber\\
    &+ \frac{\lambda}{8} \int_{p,k_1,k_2,k_3} g(-p) C_2(k_1, -k_1; k_3; p)  \sqrt{\frac{\omega_{\UV,3}}{\omega_{\Lambda,3}}} g(-k_3)  \nonumber\\
    &-\frac{\lambda}{4} \int_{p,l,k_2,k_3} \sqrt{\frac{\omega_{\UV,p}}{\omega_{\Lambda,p}}} g(-p) g(-\ell) C_{2}(p,k_2;k_3;\ell) \sqrt{\frac{\omega_{\UV,2} \omega_{\UV,3}}{\omega_{\Lambda,2} \omega_{\Lambda,3}}} g(-k_2) g(-k_3) \nonumber\\
    &
    -\frac{\lambda}{12} \int_{p,q,\ell,k} \sqrt{\frac{\omega_{\UV,q} \omega_{\UV,p}}{\omega_{\Lambda,q} \omega_{\Lambda,p}}} g(-p)g(-\ell)g(-q)C_2(p, q ; k; \ell)\sqrt{\frac{\omega_{\UV,k}}{\omega_{\Lambda,k}}}g(-k) \nonumber\\
    & + \frac{\lambda}{4} \int_{p,k} g(-p) C_3(k; p) \sqrt{\frac{\omega_{\UV,p}}{\omega_{\Lambda,p}}} g(p)  \nonumber\\
    & + \int_p g(-p) C_4(p) \sqrt{\frac{\omega_{\UV,p}}{\omega_{\Lambda,p}}} g(p)  \ ,
\end{align}
and
\begin{align}
    \label{epsilon_int}
    \Delta=&- \frac{\lambda}{4!} \int_{p,k_1,k_2,k_3} g(-p) C_{1} (k_1, k_2, k_3 ; p) \left\{ (r - r')^3 f_{0}(-k_1) f_{0}(-k_2) f_{0}(-k_3)  {\vphantom{\sqrt{\frac{\omega_{\UV,3}}{\omega_{\Lambda,3}}}}}\right.\nonumber\\
                &\quad{\hphantom{- \frac{\lambda}{4!} \int_{p,k_1,k_2,k_3} g(-p) C_{1} (k_1, k_2, k_3 ; p)}}- 3 (r - r')^2 \sqrt{\frac{\omega_{\UV,3}}{\omega_{\Lambda,3}}} f_{0}(-k_1) f_{0}(-k_2) g(-k_3) \nonumber\\
                &\quad{\hphantom{- \frac{\lambda}{4!} \int_{p,k_1,k_2,k_3} g(-p) C_{1} (k_1, k_2, k_3 ; p)}}\left. + 3 (r - r') \sqrt{\frac{\omega_{\UV,2} \omega_{\UV,3}}{\omega_{\Lambda,2} \omega_{\Lambda,3}}} f_{0}(-k_1) g(-k_2) g(-k_3)  \right\}\nonumber\\ 
                &\quad+ \frac{\lambda}{8} \int_{p,k_1,k_2} g(-p) C_{1} (k_1, k_2, -k_1 -k_2 ; p) (r - r') f_{0}(-k_2) \nonumber\\
                &\quad- \frac{\lambda}{8} \int_{p,k_1,k_2,k_3} g(-p) C_2(k_1, k_2; k_3; p) \left\{ (r - r') (r^2 + r'^2) f_{0}(-k_1) f_{0}(-k_2) f_{0}(-k_3){\vphantom{\sqrt{\frac{\omega_{\UV,1}}{\omega_{\Lambda,1}}}}} \right. \nonumber\\
                &\quad\quad{\hphantom{- \frac{\lambda}{8} \int_{p,k_1,k_2,k_3} g(-p) C_2(k_1, k_2; k_3; p)}} - 2 (r^2 - r'^2) \sqrt{\frac{\omega_{\UV,1}}{\omega_{\Lambda,1}}} g(-k_1) f_{0}(-k_2) f_{0}(-k_3) \nonumber\\
                &\quad\quad{\hphantom{- \frac{\lambda}{8} \int_{p,k_1,k_2,k_3} g(-p) C_2(k_1, k_2; k_3; p)}}
                - (r + r')^2 \sqrt{\frac{\omega_{\UV,3}}{\omega_{\Lambda,3}}} f_{0}(-k_1) f_{0}(-k_2) g(-k_3) \nonumber\\
                &\quad\quad{\hphantom{- \frac{\lambda}{8} \int_{p,k_1,k_2,k_3} g(-p) C_2(k_1, k_2; k_3; p)}} + 2 (r + r') \sqrt{\frac{\omega_{\UV,2} \omega_{\UV,3}}{\omega_{\Lambda,2} \omega_{\Lambda,3}}} f_{0}(-k_1)g(-k_2) g(-k_3) \nonumber\\
                &\quad\quad\left.{\hphantom{- \frac{\lambda}{8} \int_{p,k_1,k_2,k_3} g(-p) C_2(k_1, k_2; k_3; p)}}
                + (r - r') \sqrt{\frac{\omega_{\UV,1} \omega_{\UV,2}}{\omega_{\Lambda,1} \omega_{\Lambda,2}}} g(-k_1) g(-k_2) f_{0}(-k_3)  \right\}   \nonumber\\
                &\quad + \frac{\lambda}{4} \int_{p,k_1,k_2,k_3} g(-p) C_2(k_1, k_2; -k_1; p) (r + r') f_{0}(-k_2) \nonumber\\
                &\quad - \frac{\lambda}{8} \int_{p,k_1,k_2,k_3} g(-p) C_2(k_1, -k_1; k_3; p) (r - r') f_{0}(-k_3)\nonumber\\
                &\quad-\frac{\lambda}{4} \int_{p,l,k_2,k_3} \sqrt{\frac{\omega_{\UV,p}}{\omega_{\Lambda,p}}} g(-p) g(-\ell) C_{2}(p,k_2;k_3;\ell) \nonumber\\
                &\quad\qquad\qquad \times \left\{ \qty( r^2 - r'^2 ) f_{0}(-k_2) f_{0}(-k_3) - (r + r') \sqrt{\frac{\omega_{\UV,3}}{\omega_{\Lambda,3}}} f_{0}(-k_2) g(-k_3) \right. \nonumber\\
                &\quad\qquad\qquad \qquad{\hphantom{{\qty( r^2 - r'^2 ) f_{0}(-k_2) f_{0}(-k_3)}}}- (r - r') \sqrt{\frac{\omega_{\UV,2}}{\omega_{\Lambda,2}}} f_{0}(-k_3) g(-k_2)  \Bigg\}\nonumber\\
                &\quad+\frac{\lambda}{12} \int_{p,q,\ell,k} \sqrt{\frac{\omega_{\UV,q} \omega_{\UV,p}}{\omega_{\Lambda,q} \omega_{\Lambda,p}}} g(-p)g(-\ell)g(-q)C_2(p, q ; k; \ell) \qty(r-r')f_{0}(-k) \nonumber\\
                &\quad- \frac{\lambda}{4} \int_{p,k} g(-p) C_3(k; p) \qty(r+r') f_{0}(p) \nonumber\\
                &\quad - \int_p g(-p) C_4(p) \qty(r-r') f_{0}(p) \ .
\end{align}
Note that $\Gamma$, which is a contribution 
from the interactions, is independent of $r$ and $r'$.
When
$\varepsilon_{\text{free},\Lambda}$ and $\alpha\varepsilon_{\text{int},\Lambda}$,
which are dependent on $r$ and $r'$, are small,
the Knill-Laflamme condition is satisfied in an approximate
sense.
We see that $M[g,h]$ in \eqref{KL condition as exp value} is different from that in the free case. 
We have already found the situations where $\varepsilon_{\text{free},\Lambda}$ is small.
Unlike $\varepsilon_{\text{free},\Lambda}$,
$\alpha\varepsilon_{\text{int},\Lambda}$ is not small
even when an approximately orthogonal basis is used.
The smaller the combination of the strength of
perturbation, the overlap between $f_0$ and $g$ and
the energy scale $\Lambda/m$, the smaller 
$\alpha\varepsilon_{\text{int},\Lambda}$ is.
It is easy to see that $C_1,\,C_2$, and $C_3$ behave as $(\Lambda/m)^{1/2}$ which comes from the second term of each.
On the other hand, we need to be careful with $C_4$. It has $\delta m^2_{\Lambda}$ in first term, which behaves as $(\Lambda/m)^{-2}$, as detailed in Appendix E.
This implies that $C_4$ also behaves as $(\Lambda/m)^{1/2}$.
Thus, $\alpha\varepsilon_{\text{int},\Lambda}$ behaves as
$(\Lambda/m)^{1/2}$ in the same way as
$\varepsilon_{\text{free},\Lambda}$.

We therefore found the situations where
the code subspace spanned by $\{\Ket{rf_{0}}\}$ 
is error correctable for $D[g]$
to the first order in
the perturbation theory.
The combination of 
the orthogonality of the basis, the strength of perturbation,
the locality of $g$ and the energy scale $\Lambda/m$
determines how accurately the Knill-Laflamme condition is
satisfied.

\section{Conclusion and discussion}
In this paper, we demonstrated that quantum error correction 
is realized by the renormalization group
in scalar field theories. 
We constructed $q$-level states locally 
by using coherent states in the IR region. 
By acting on them the inverse of the unitary operator $U$ that describes the 
renormalization group flow of the ground state, we encoded them into
states in the UV region. 
We examined whether the Knill-Laflamme condition is satisfied
for a class of error operators that have the same form as the operators that create coherent states, 
to the first order
in the perturbation theory.
We estimated the deviation from the exact
Knill-Laflamme condition.
We found that the combination of
the orthogonality of the basis, the strength of perturbation,
the locality of the error and the energy scale
determines how accurately 
the Knill-Laflamme condition is satisfied.



Some future directions are in order.
We would like to extend the analysis in this paper 
to the non-perturbative one, where the exact renormalization group equation for
wave functionals derived in \cite{Kuwahara:2022nlm} is expected to be useful.
It is interesting to consider other realizations of $q$-level systems in 
the IR region or other types of error operators. In those cases, other types of the inverse of the 
renormalization group may be needed. Indeed, taking the inverse
of the renormalization group 
\cite{Berman:2022uov, Berman:2023rqb,Cotler:2023lem} is nontrivial, 
because the renormalization group is a semi-group.
The above directions would be relevant for
revealing a general relationship between the renormalization group and 
quantum error correction.
As well as extension to the non-perturbative analysis, extension
to the cases of gauge theories is important from the viewpoint of 
the gauge/gravity correspondence, because the strongly coupled
regime at large $N$ in the boundary theory corresponds to 
classical geometry in the bulk.
We hope to report developments in these directions in the near future.

\section*{Acknowledgments}
We would like to thank Kazushi Yamashiro for collaboration in the early stage of this work.
Useful discussions with the participants of the workshop 
“Large-N matrix models and emergent geometry” 
at the Erwin-Schr\"odinger Institute for Mathematics and Physics 
in Vienna 
and Workshop on Noncommutative and Generalized Geometry 
in String Theory, Gauge Theory and Related Physical Models in Corfu 
are acknowledged.
G.T. was supported by JSPS KAKENHI Grant Numbers JP22H01222.
A.T. was supported in part by JSPS KAKENHI Grant Numbers JP21K03532 and
by MEXT KAKENHI Grant in Aid for Transformative
Research Areas A ``Extreme Universe” No. 22H05253.

\section*{Appendix A: Calculation of $\Psi_{\Lambda}^{(1)}$}
\renewcommand{\theequation}{A.\arabic{equation}}
\setcounter{equation}{0}
In this section, we calculate the first-order wave functional $\Psi_\Lambda^{(1)}$ in the perturbation theory\cite{hatfield2018quantum}.
$\Psi_\Lambda^{(1)}$ is given in terms of the zeroth order wave functional $\Psi_\Lambda^{(0)}$ as follows:
\begin{align}
	\Psi_0^{(1)} &= \sum_{n \neq 0} \int_{k_1,k_2,\dots,k_n}
		\frac{ \Braket{ \Psi_n^{(0)}(k_1,k_2,\dots,k_n)|H_\mathrm{int} | \Psi_\Lambda^{(0)} } }{E^{(0)}_0-E^{(0)}_n(k_1,k_2,\dots, k_n)}
		\Psi_n^{(0)}(k_1,k_2,\dots,k_n) \ ,	\label{1storderPsi}
\end{align}
where
\begin{align}
	\Psi_n^{(0)}(k_1,k_2,\dots,k_n) = \frac{1}{\sqrt{n!}} \prod_{i=1}^n a_{\Lambda, k_i}^{(0)\dagger} \Psi_\Lambda^{(0)}
\end{align}
is the zeroth-order $n$-particle wave functional in the perturbation theory,
and $E^{(0)}_n(k_i)$ is its energy eigenvalue.

To obtain $\Psi_\Lambda^{(1)}$, we calculate the RHS of ({\ref{1storderPsi}) with $n=2$ and 4 as we consider the $\phi^4$ theory.
First, in the case of $n=4$, we have
\begin{align}
	&\int_{k_1,k_2,k_3,k_4}
		\frac{ \Braket{ \Psi_4^{(0)}(k_1,k_2,k_3,k_4)|H_\mathrm{int} | \Psi_\Lambda^{(0)} } }{E^{(0)}_0-E^{(0)}_4(k_1,k_2,k_3,k_4)}
		\Psi_4^{(0)}(k_1,k_2,k_3,k_4)	\nonumber	\\
	=& \frac{\lambda}{4!} \int_{k_1,\dots,k_4}
		\frac{ \Braket{ \Psi_4^{(0)}(k_1,\dots,k_4)
		| \int_{p_i} \phi(p_1) \phi(p_2) \phi(p_3) \phi(p_4) \tilde{\delta}(\sum_{i=1}^4p_i)
		| \Psi_\Lambda^{(0)} } }{E^{(0)}_0-E^{(0)}_4(k_1,\dots,k_4)} \Psi_4^{(0)}(k_1,\dots,k_4)
		\nonumber	\\
	=& \frac{\lambda}{4!} \int_{k_1,\dots,k_4}
		\frac{ \Braket{ \Psi_4^{(0)}(k_1,\dots,k_4)
		| \int_{p_i} \prod_{m=1}^4 \sqrt{\frac{K_{p_m}}{2\omega_{\Lambda, p_m}}} a_{\Lambda, p_m}^{(0)\dagger} \tilde{\delta}(\sum_{i=1}^4p_i)
		| \Psi_\Lambda^{(0)} } }{E^{(0)}_0-E^{(0)}_4(k_1,\dots,k_4)} \Psi_4^{(0)}(k_1,\dots,k_4)
		\nonumber	\\
	=& \frac{\lambda}{4!} \int_{k_1,\dots,k_4}
		\frac{\sqrt{4!} \tilde{\delta}(\sum_{i=1}^4 k_i)}{- \omega_{\Lambda,k_1} - \omega_{\Lambda,k_2} - \omega_{\Lambda,k_3} - \omega_{\Lambda,k_4}}
		\prod_{i=1}^4 \sqrt{\frac{K_{k_i}}{2\omega_{\Lambda, k_i}}}
		\Psi_4^{(0)}(k_1,\dots,k_4)
		\nonumber	\\
	=& - \frac{\lambda}{4!} \int_{k_1,\dots,k_4}
		\frac{ \tilde{\delta}(k_1+k_2+k_3+k_4)}{\omega_{\Lambda,k_1} + \omega_{\Lambda,k_2} + \omega_{\Lambda,k_3} + \omega_{\Lambda,k_4}}
		\prod_{i=1}^4 \sqrt{\frac{K_{k_i}}{2\omega_{\Lambda, k_i}}} a_{\Lambda, k_i}^{(0)\dagger}
		\Psi_\Lambda^{(0)}	\ .
\end{align}
Next, in the case of $n=2$, we have
\begin{align}
	&\int_{k_1,k_2}
		\frac{ \Braket{ \Psi_2^{(0)}(k_1,k_2)|H_\mathrm{int} | \Psi_\Lambda^{(0)} } }{E^{(0)}_0-E^{(0)}_2(k_1,k_2)}
		\Psi_2^{(0)}(k_1,k_2)	\nonumber	\\
	=& \frac{\lambda}{4!} \int_{k_1,k_2}
		\frac{ \Braket{ \Psi_2^{(0)}(k_1,k_2)
		| \int_{p_i} \phi(p_1) \phi(p_2) \phi(p_3) \phi(p_4) \tilde{\delta}(\sum_{i=1}^4p_i)
		| \Psi_\Lambda^{(0)} } }{E^{(0)}_0-E^{(0)}_2(k_1,k_2)} \Psi_2^{(0)}(k_1,k_2)
		\nonumber	\\
	&+ \frac{\delta m^2}{2} \int_{k_1,k_2}
		\frac{ \Braket{ \Psi_2^{(0)}(k_1,k_2)
		| \int_{p} \phi(p) \phi(-p)
		| \Psi_\Lambda^{(0)} } }{E^{(0)}_0-E^{(0)}_2(k_1,k_2)} \Psi_2^{(0)}(k_1,k_2)
		\nonumber	\\
	=& \frac{\lambda}{4!} \int_{k_1,k_2}
		\frac{ \Braket{ \Psi_2^{(0)}(k_1,k_2)
		| \int_{p_i} 6\sqrt{\frac{K_{p_1}}{2\omega_{\Lambda, p_1}}} a_{\Lambda, -p_1}^{(0)} 
			\prod_{m=2}^4 \sqrt{\frac{K_{p_m}}{2\omega_{\Lambda, p_m}}} a_{\Lambda, p_m}^{(0)\dagger} \tilde{\delta}(\sum_{i=1}^4p_i)
		| \Psi_\Lambda^{(0)} } }{E^{(0)}_0-E^{(0)}_2(k_1,k_2)} \Psi_2^{(0)}(k_1,k_2)
		\nonumber	\\
	&+ \frac{\delta m^2}{2} \int_{k_1,k_2}
		\frac{ \Braket{ \Psi_2^{(0)}(k_1,k_2)
		| \int_{p} \frac{K_p}{2\omega_{\Lambda, p}} a_{\Lambda, -p}^{(0)\dagger} a_{\Lambda, p}^{(0)\dagger}
		| \Psi_\Lambda^{(0)} } }{E^{(0)}_0-E^{(0)}_2(k_1,k_2)} \Psi_2^{(0)}(k_1,k_2)
		\nonumber	\\
	=& \frac{\lambda}{4!} \int_{k_1,k_2}
		\frac{ \sqrt{2}\tilde{\delta}(k_1+k_2)}{- \omega_{\Lambda,k_1} - \omega_{\Lambda,k_2}} \int_p \frac{6K_p}{2\omega_{\Lambda,p}}
		\Psi_2^{(0)}(k_1,k_2)	
		+ \frac{\delta m^2}{2} \int_{k_1,k_2} \frac{ \sqrt{2}\tilde{\delta}(k_1+k_2)}{- \omega_{\Lambda,k_1} - \omega_{\Lambda,k_2}}
		\Psi_2^{(0)}(k_1,k_2)		\nonumber	\\
	=& - \left( \frac{\delta m^2}{2} + \frac{\lambda}{4!} \int_p \frac{6K_p}{2\omega_{\Lambda,p}} \right)
		\int_k \frac{1}{2\omega_{\Lambda,k}}
		\frac{K_k}{2\omega_{\Lambda,k}} {a}^{(0)\dagger}_{\Lambda,k} {a}^{(0)\dagger}_{\Lambda,-k} \Psi_\Lambda^{(0)} \ .
\end{align}

Finally, combining these results, we obtain
\begin{align}
	\Psi_\Lambda^{(1)}
	=& \left[ \left\{ - \frac{\lambda}{4!} \int_{k_1,k_2,k_3,k_4}
		\frac{ \tilde{\delta}(k_1+k_2+k_3+k_4)}{\omega_{\Lambda,k_1} + \omega_{\Lambda,k_2} + \omega_{\Lambda,k_3} + \omega_{\Lambda,k_4}}
		\prod_{i=1}^4 \sqrt{\frac{K_{k_i}}{2\omega_{\Lambda, k_i}}} a_{\Lambda, k_i}^{(0)\dagger} \right. \right.	\nonumber	\\
	&\left. \left. - \left( \frac{\delta m^2}{2} + \frac{\lambda}{4!} \int_p \frac{6K_p}{2\omega_{\Lambda,p}} \right)
		\int_k \frac{1}{2\omega_{\Lambda,k}}
		\frac{K_k}{2\omega_{\Lambda,k}} {a}^{(0)\dagger}_{\Lambda,k} {a}^{(0)\dagger}_{\Lambda,-k} \right\} - \{\mathrm{Hermitian\ conjugate}\} \right] 
		\Psi_\Lambda^{(0)}	\ ,
\end{align}
where we used $a_{\Lambda,k}^{(0)} \Psi_{\Lambda}^{(0)}=0$. 
This result agrees with the one obtained in \cite{Kuwahara:2022nlm} using the path integral.

Using this result, we can write $\Psi_\Lambda^{(1)}$ as
\begin{align}
	\Psi^{(1)}_{\Lambda} = A_{\Lambda} \Psi^{(0)}_{\Lambda} \ ,
\end{align}
where
\begin{align}
	A_{\Lambda}
	=& \left\{ - \frac{\lambda}{4!} \int_{k_1,k_2,k_3,k_4}
		\frac{ \tilde{\delta}(k_1+k_2+k_3+k_4)}{\omega_{\Lambda,k_1} + \omega_{\Lambda,k_2} + \omega_{\Lambda,k_3} + \omega_{\Lambda,k_4}}
		\prod_{i=1}^4 \sqrt{\frac{K_{k_i}}{2\omega_{\Lambda, k_i}}} a_{\Lambda, k_i}^{(0)\dagger} \right.	\nonumber	\\
	&\left. - \left( \frac{\delta m^2}{2} + \frac{\lambda}{4!} \int_p \frac{6K_p}{2\omega_{\Lambda,p}} \right)
		\int_k \frac{1}{2\omega_{\Lambda,k}}
		\frac{K_k}{2\omega_{\Lambda,k}} {a}^{(0)\dagger}_{\Lambda,k} {a}^{(0)\dagger}_{\Lambda,-k} \right\} - \{\mathrm{Hermitian\ conjugate}\}
		\ .
\end{align}

\section*{Appendix B: The explicit form of $X_{\Lambda}^{(1)}$}
\renewcommand{\theequation}{B.\arabic{equation}}
\setcounter{equation}{0}
In this section, we determine $X_\Lambda^{(1)}$ using (\ref{flow equation for A}):
\begin{align}
     -\Lambda\partial_{\Lambda}A_{\Lambda} 		= X^{(1)}_{\Lambda}+[X^{(0)}_{\Lambda},A_{\Lambda}] \ .
\end{align}
See \cite{Cotler:2018ufx} for a similar perturbative calculation of $X_\Lambda$.

First, we calculate $-\Lambda\partial_{\Lambda}A_{\Lambda}$. Using the explicit form of $A_\Lambda$ obtained in the last section, we have
\begin{align}
    &-\Lambda \partial_\Lambda A_\Lambda \Psi_\Lambda^{(0)}   \nonumber   \\
    &= -\Lambda \partial_\Lambda \left\{ - \frac{\lambda}{4!} \int_{k_1,k_2,k_3,k_4}
        \frac{ \tilde{\delta}(k_1+k_2+k_3+k_4)}{\omega_{\Lambda,k_1} + \omega_{\Lambda,k_2} + \omega_{\Lambda,k_3} + \omega_{\Lambda,k_4}}
        \prod_{i=1}^4 \sqrt{\frac{K_{k_i}}{2\omega_{\Lambda, k_i}}} a_{\Lambda, k_i}^{(0)\dagger} \right.	\nonumber	\\
    &\hspace{70pt} \left. - \left( \frac{\delta m^2}{2} + \frac{\lambda}{4!} \int_p \frac{6K_p}{2\omega_{\Lambda,p}} \right)
        \int_k \frac{1}{2\omega_{\Lambda,k}}
        \frac{K_k}{2\omega_{\Lambda,k}} {a}^{(0)\dagger}_{\Lambda,k} {a}^{(0)\dagger}_{\Lambda,-k} \right\} - \{ \mathrm{Hermitian\ conjugate}\} \nonumber   \\
    &= \left\{ \frac{\lambda}{4!} \int_{k_1,k_2,k_3,k_4}
        \frac{\dot{\omega}_{\Lambda,k_1} + \dot{\omega}_{\Lambda,k_2} + \dot{\omega}_{\Lambda,k_3} + \dot{\omega}_{\Lambda,k_4}}
            {(\omega_{\Lambda,k_1} + \omega_{\Lambda,k_2} + \omega_{\Lambda,k_3} + \omega_{\Lambda,k_4})^2}
        \tilde{\delta}(k_1+k_2+k_3+k_4)
        \prod_{i=1}^4 \sqrt{\frac{K_{k_i}}{2\omega_{\Lambda, k_i}}} a_{\Lambda, k_i}^{(0)\dagger} \right.	\nonumber	\\
    &\hspace{20pt} + \frac{\lambda}{4!} \int_{k_1,k_2,k_3,k_4}
        \frac{ \tilde{\delta}(k_1+k_2+k_3+k_4)}{\omega_{\Lambda,k_1} + \omega_{\Lambda,k_2} + \omega_{\Lambda,k_3} + \omega_{\Lambda,k_4}}
        \left( \sum_{i=1}^4 \frac{\dot{\omega}_{\Lambda, k_i}}{2\omega_{\Lambda, k_i}} \right)
        \prod_{i=1}^4 \sqrt{\frac{K_{k_i}}{2\omega_{\Lambda, k_i}}} a_{\Lambda, k_i}^{(0)\dagger}   \nonumber   \\
    &\hspace{20pt} - \frac{\lambda}{4!} \int_{k_1,k_2,k_3,k_4}
        \frac{ \tilde{\delta}(k_1+k_2+k_3+k_4)}{\omega_{\Lambda,k_1} + \omega_{\Lambda,k_2} + \omega_{\Lambda,k_3} + \omega_{\Lambda,k_4}}
        \left( \prod_{i=1}^4 \sqrt{\frac{K_{k_i}}{2\omega_{\Lambda, k_i}}} \right) \frac{4\dot{\omega}_{\Lambda, k_4}}{2\omega_{\Lambda, k_4}}
        a_{\Lambda, k_1}^{(0)\dagger} a_{\Lambda, k_2}^{(0)\dagger} a_{\Lambda, k_3}^{(0)\dagger} a_{\Lambda, -k_4}^{(0)}
        \nonumber\\
    &\hspace{20pt} - \frac{\lambda}{4!} \int_{k_1,k_2}
        \frac{1}{2\omega_{\Lambda,k_1} + 2\omega_{\Lambda,k_2}}
        \frac{K_{k_2}}{2\omega_{\Lambda, k_2}} \frac{K_{k_2}}{2\omega_{\Lambda, k_2}} \frac{6\dot{\omega}_{\Lambda, k_2}}{2\omega_{\Lambda, k_2}}
        a_{\Lambda, k_1}^{(0)\dagger} a_{\Lambda, -k_1}^{(0)\dagger}   \nonumber	\\
	&\hspace{20pt} - \left( \frac{\dot{\delta m^2}}{2} - \frac{\lambda}{4!} \int_p \frac{6K_p \dot{\omega}_{\Lambda,p}}{2\omega^2_{\Lambda,p}} \right)
		\int_k \frac{1}{2\omega_{\Lambda,k}}
		\frac{K_k}{2\omega_{\Lambda,k}} {a}^{(0)\dagger}_{\Lambda,k} {a}^{(0)\dagger}_{\Lambda,-k}	\nonumber\\
	&\hspace{20pt} \left. + \left( \frac{\delta m^2}{2} + \frac{\lambda}{4!} \int_p \frac{6K_p}{2\omega_{\Lambda,p}} \right)
		\int_k \frac{K_k}{2\omega^3_{\Lambda,k}} {a}^{(0)\dagger}_{\Lambda,k} {a}^{(0)\dagger}_{\Lambda,-k} \right\}
		- \left\{\mathrm{Hermitian\ conjugate} \right\}	\ .
\end{align}

Next, we calculate $[X^{(0)}_{\Lambda},A_{\Lambda}]$. Substituting (\ref{X^(0)}) and (\ref{A}) into the commutator, we obtain
\begin{align}
	&[X^{(0)}_{\Lambda},A_{\Lambda}]	\nonumber	\\
	&= \left[ -\frac{1}{4} \int_p \frac{\dot{\omega}_{\Lambda, p}}{\omega_{\Lambda, p}} 
		\left( a^{(0)\dagger}_{\Lambda, p} a^{(0)\dagger}_{\Lambda, -p}	- a^{(0)}_{\Lambda, p} a^{(0)}_{\Lambda, -p} \right) ,\right. \nonumber	\\
	&\hspace{30pt} - \frac{\lambda}{4!} \int_{k_1,k_2,k_3,k_4}
		\frac{ \tilde{\delta}(k_1+k_2+k_3+k_4)}{\omega_{\Lambda,k_1} + \omega_{\Lambda,k_2} + \omega_{\Lambda,k_3} + \omega_{\Lambda,k_4}}
		\left( \prod_{i=1}^4 \sqrt{\frac{K_{k_i}}{2\omega_{\Lambda, k_i}}} a_{\Lambda, k_i}^{(0)\dagger}
		- \prod_{i=1}^4 \sqrt{\frac{K_{k_i}}{2\omega_{\Lambda, k_i}}} a_{\Lambda, k_i}^{(0)} \right)	\nonumber	\\
	&\hspace{50pt}\left. - \left( \frac{\delta m^2}{2} + \frac{\lambda}{4!} \int_p \frac{6K_p}{2\omega_{\Lambda,p}} \right)
		\int_k \frac{1}{2\omega_{\Lambda,k}}
		\frac{K_k}{2\omega_{\Lambda,k}}
		\left({a}^{(0)\dagger}_{\Lambda,k} {a}^{(0)\dagger}_{\Lambda,-k} - {a}^{(0)}_{\Lambda,k} {a}^{(0)}_{\Lambda,-k} \right) \right]
		\nonumber\\
	&= \left\{ - \frac{\lambda}{4!}\int_{k_1,k_2} \frac{1}{2\omega_{\Lambda, k_1} + 2\omega_{\Lambda, k_2}}
		\frac{K_{k_1}}{2\omega_{\Lambda, k_1}} \frac{K_{k_2}}{2\omega_{\Lambda, k_2}} \frac{3\dot{\omega}_{\Lambda, k_2}}{\omega_{\Lambda, k_2}}
		a_{\Lambda, k_1}^{(0)\dagger} a_{\Lambda, -k_1}^{(0)\dagger} \right.	\nonumber	\\
	&\hspace{10pt} \left. - \frac{\lambda}{4!} \int_{k_1, k_2, k_3, k_4} \frac{1}{2\omega_{\Lambda, k_1} + 2\omega_{\Lambda, k_2}}
		\frac{K_{k_1}}{2\omega_{\Lambda, k_1}} \frac{K_{k_2}}{2\omega_{\Lambda, k_2}} \frac{2\dot{\omega}_{\Lambda, k_2}}{\omega_{\Lambda, k_2}}
		a^{(0)\dagger}_{\Lambda, k_1} a^{(0)\dagger}_{\Lambda, k_2} a^{(0)\dagger}_{\Lambda, k_3} a^{(0)}_{\Lambda, -k_4} \right\}    \nonumber\\
	&\hspace{10pt}- \left\{ \mathrm{Hermitian \ conjugate} \right\} \ .
\end{align}

Combining these results, we have
\begin{align}
	X^{(1)}_\Lambda
	&= \left\{ \frac{\lambda}{4!} \int_{k_1,k_2,k_3,k_4}
		\frac{\tilde{\delta}(k_1+k_2+k_3+k_4)}
		{(\omega_{\Lambda,k_1} + \omega_{\Lambda,k_2} + \omega_{\Lambda,k_3} + \omega_{\Lambda,k_4})^2}
		\left( \sum_{i=1}^4 \dot{\omega}_{\Lambda, k_i} \right)
		\prod_{i=1}^4 \sqrt{\frac{K_{k_i}}{2\omega_{\Lambda, k_i}}} a_{\Lambda, k_i}^{(0)\dagger} \right.	\nonumber	\\
	&\hspace{20pt} + \frac{\lambda}{4!} \int_{k_1,k_2,k_3,k_4}
		\frac{ \tilde{\delta}(k_1+k_2+k_3+k_4)}{\omega_{\Lambda,k_1} + \omega_{\Lambda,k_2} + \omega_{\Lambda,k_3} + \omega_{\Lambda,k_4}}
		\left( \sum_{i=1}^4 \frac{\dot{\omega}_{\Lambda, k_i}}{2\omega_{\Lambda, k_i}} \right)
		\prod_{i=1}^4 \sqrt{\frac{K_{k_i}}{2\omega_{\Lambda, k_i}}} a_{\Lambda, k_i}^{(0)\dagger}   \nonumber   \\
	&\hspace{20pt} - \left( \frac{\dot{\delta m^2}}{2} - \frac{\lambda}{4!} \int_p \frac{6K_p \dot{\omega}_{\Lambda,p}}{2\omega^2_{\Lambda,p}} \right)
		\int_k \frac{1}{2\omega_{\Lambda,k}}
		\frac{K_k}{2\omega_{\Lambda,k}} {a}^{(0)\dagger}_{\Lambda,k} {a}^{(0)\dagger}_{\Lambda,-k}	\nonumber\\
	&\hspace{20pt} \left. + \left( \frac{\delta m^2}{2} + \frac{\lambda}{4!} \int_p \frac{6K_p}{2\omega_{\Lambda,p}} \right)
		\int_k \frac{K_k}{2\omega^3_{\Lambda,k}} {a}^{(0)\dagger}_{\Lambda,k} {a}^{(0)\dagger}_{\Lambda,-k}	\right\}
		- \left\{ \mathrm{Hermitian \ conjugate} \right\} \ .
\end{align}
This is an anti-Hermitian operator as required.

\section*{Appendix C: Calculation of $D[g]$}
\renewcommand{\theequation}{C.\arabic{equation}}
\setcounter{equation}{0}
In this section, we calculate $D[g]$ up to the first order of perturbation in the interacting case.
$D[g]$ is given as follows:
\begin{align}
D[g] &= \exp \left[ \int_{p} g(-p) a^{-}_{\Lambda,p} \right] = \exp \left[ \int_{p} g(-p) ( a^{-(0)}_{\Lambda,p} + \alpha a^{-(1)}_{\Lambda,p} ) \right].
\end{align}
Here, the contents of the exponent can symbolically be written as follows:
\begin{align}
    a_{-,\Lambda}(p)
    &=a^{-(0)}_{\Lambda}(p)+\alpha a^{-(1)}_{\Lambda,p}\qty(a^{+(0)}_{\Lambda},a^{-(0)}_{\Lambda})\\
    &=\sqrt{\frac{\omega_{\UV,p}}{\omega_{\Lambda,p}}}a^{-(0)}_{\UV,p}+\alpha a^{-(1)}_{\Lambda,p}\qty(\sqrt{\frac{\omega_{\Lambda}}{\omega_{\UV}}}a^{+(0)}_{\UV},\sqrt{\frac{\omega_{\UV}}{\omega_{\Lambda}}}a^{-(0)}_{\UV}).
\end{align}
This notation indicates that $a^{-(1)}_{\Lambda}$ includes $a^{\pm(0)}_{\UV}$.
Next, in the first term, we use $a^{-(0)}_{\UV}=a^{-}_{\UV}-\alpha a^{-(1)}_{\UV}$. 
Then, in the term proportional to $\alpha$ , it is allowed to replace $a^{(0)}$ with $a$ since we consider up to the first order of perturbation. 
Then, we obtain
\begin{align}
    a^{-}_{\Lambda,p}
   =\sqrt{\frac{\omega_{\UV,p}}{\omega_{\Lambda,p}}}a^{-}_{\UV,p}+\alpha \qty{a^{-(1)}_{\Lambda,p}\qty(\sqrt{\frac{\omega_{\Lambda}}{\omega_{\UV}}}a^{+}_{\UV},\sqrt{\frac{\omega_{\UV}}{\omega_{\Lambda}}}a^{-}_{\UV})-\sqrt{\frac{\omega_{\UV,p}}{\omega_{\Lambda,p}}}a^{-(1)}_{\UV,p}\qty(a^{+}_{\UV},a^{-}_{\UV})}.\label{a^-}
\end{align}

On the other hand, we calculate the concrete form of $a^{-(1)}_{\UV,p}$ from \eqref{solution of a1 scaling} as follows:
\begin{align}
a^{-(1)}_{\Lambda}(p) 
&= a^{(1)}_{\Lambda}(p) - a^{(1)\dag}_{\Lambda}(-p) \notag\\
&\begin{aligned}
    &= -\frac{\lambda}{3!} \int_{k_{1},k_{2},k_{3}}\frac{\tilde{\delta}(k_{4} + \dots + p)}{\omega_{\Lambda,1}+\cdots+\omega_{\Lambda,p}}\qty(\prod_{i=1}^{3} \sqrt{\frac{K_{i}}{2\omega_{i}}})\sqrt{\frac{K_{p}}{2\omega_{p}}}\\
    &\quad\quad\times\qty(a^{(0)}_{\Lambda,-k_1} a^{(0)}_{\Lambda,-k_2} a^{(0)}_{\Lambda,-k_3} - a^{(0)\dag}_{\Lambda,k_1} a^{(0)\dag}_{\Lambda,k_2} a^{(0)\dag}_{\Lambda,k_3}) \\
    &\quad -2 \qty( \frac{\delta m^2}{2} + \frac{\lambda}{4!}\int_{q}\frac{6K_{p}}{2\omega_{\Lambda,q}})\frac{K_{p}}{4\omega^2_{\Lambda,p}} \qty(a^{(0)}_{\Lambda,p} - a^{(0)\dag}_{\Lambda,-p}).
\end{aligned}\label{intermediate1}
\end{align}
Next, in the first term, by rewriting the creation and annihilation operators in terms of $a^{\pm(0)}_{\Lambda}$ and bringing $a^{+(0)}$ to the left and $a^{-(0)}$ to the right, we obtain
\begin{align}
&a^{(0)}_{\Lambda,-k_1} a^{(0)}_{\Lambda,-k_2} a^{(0)}_{\Lambda,-k_3} - a^{(0)\dagger}_{\Lambda,k_1} a^{(0)\dagger}_{\Lambda,k_2} a^{(0)\dagger}_{\Lambda,k_3} \notag\\
&\rightarrow \frac{1}{4} \left( a^{-(0)}_{\Lambda,-k_1} a^{-(0)}_{\Lambda,-k_2} a^{-(0)}_{\Lambda,-k_3} + 3 a^{+(0)}_{\Lambda,-k_1} a^{+(0)}_{\Lambda,-k_2} a^{-(0)}_{\Lambda,-k_3} + 3 \cdot 2 a^{+(0)}_{\Lambda,-k_1} \tilde{\delta}(-k_3-k_2) \right).
\end{align}
Here, we use the fact that the first term in (\ref{intermediate1})  is symmetric with respect to the permutation of $k_1,\,k_2$, and $k_3$. 
Then, we obtain
\begin{align}
\begin{aligned}
a^{-(1)}_{\Lambda,p} &= -\frac{\lambda}{4!} \int_{k_1k_2k_3}\frac{\tilde{\delta}(k_{1} + \dots + p)}{\omega_{\Lambda,1}+\cdots+\omega_{\Lambda,p}}\qty(\prod_{i=1}^{3} \sqrt{\frac{K_i}{2\omega_{\Lambda,i}}})\sqrt{\frac{K_p}{2\omega_{\Lambda,p}}}a^{-(0)}_{\Lambda,-k_1} a^{-(0)}_{\Lambda,-k_2} a^{-(0)}_{\Lambda,-k_3} \\
&\quad\quad-\frac{\lambda}{8} \int_{k_1,k_2,k_3}\frac{\tilde{\delta}(k_{1} + \dots + p)}{\omega_{\Lambda,1}+\cdots+\omega_{\Lambda,p}}\qty(\prod_{i=1}^{3} \sqrt{\frac{K_i}{2\omega_{\Lambda,i}}})\sqrt{\frac{K_p}{2\omega_{\Lambda,p}}}a^{+(0)}_{\Lambda,-k_1} a^{-(0)}_{\Lambda,-k_2} a^{-(0)}_{\Lambda,-k_3} \\
&\quad\quad -\frac{\lambda}{4} \int_{k} \frac{1}{2\omega_{\Lambda,p} + 2\omega_{\Lambda,k}} \frac{K_p}{2\omega_{\Lambda,p}} \frac{K_k}{2\omega_{\Lambda,k}}a^{+(0)}_{\Lambda,-k} \\
&\quad\quad - \qty( \frac{\delta m^2}{2} + \frac{\lambda}{4!} \int_{q}\frac{6 K_{q}}{2\omega_{\Lambda, q}})\frac{K_p}{2\omega^2_{\Lambda,p}} a^{-(0)}_{\Lambda,p} \ .
\end{aligned}\label{concrete_form_of_a^(1)}
\end{align}
From (\ref{a^-}) and (\ref{concrete_form_of_a^(1)}), we obtain (\ref{perturbative_error_op}).

\section*{Appendix D: Calculation of (\ref{error exp value 2}) }
\renewcommand{\theequation}{D.\arabic{equation}}
\setcounter{equation}{0}
In this section, we calculate (\ref{error exp value 2}) explicitly. 
The zeroth-order term in $\alpha$ is the same as in the free case.
Then, we evaluate the first-order term in $\alpha$.
The $[X,[X,Y]]$ term is
\begin{align}
&\sideset{_{\UV}}{_{\UV}}{\mathop{\mel**{r'f_{0}}{\left( - \frac{1}{12} [X,[X,Y]] \right)}{rf_{0}-\sqrt{\frac{\omega_{\UV}}{\omega_{\Lambda}}}g}}}\notag\\
&\begin{aligned}
&= - \frac{1}{12} \left( -\lambda  \int_{p,q,\ell,k} \sqrt{\frac{\omega_{\UV,q} \omega_{\UV,p}}{\omega_{\Lambda,q} \omega_{\Lambda,p}}} g(-p)g(-\ell)g(-q) C_2(p, q ; k; \ell) \right)\\
&\quad\quad\quad\quad \times  \sideset{_{\UV}}{_{\UV}}{\mathop{\mel**{r'f_{0}}{\qty( a_{\UV,-k} - a_{\UV, k}^\dagger )}{rf_{0}-\sqrt{\frac{\omega_{\UV}}{\omega_{\Lambda}}}g}}} \\
\end{aligned} \nonumber \\
&\begin{aligned}
&= \frac{\lambda}{12} \int_{p,q,\ell,k} \sqrt{\frac{\omega_{\UV,q} \omega_{\UV,p}}{\omega_{\Lambda,q} \omega_{\Lambda,p}}} g(-p)g(-\ell)g(-q) C_2(p, q ; k; \ell) \\
&\quad\quad\quad\quad \times \left\{ \qty(r-r')f_{0}(-k) - \sqrt{\frac{\omega_{\UV,k}}{\omega_{\Lambda,k}}}g(-k) \right\}\sideset{_{\UV}}{_{\UV}}{\mathop{\Braket{r'f_{0}|rf_{0}-\sqrt{\frac{\omega_{\UV}}{\omega_{\Lambda}}}g}}}.
\end{aligned}
\end{align}
The $[X,Y]$ term is
\begin{align}
&\sideset{_{\UV}}{_{\UV}}{\mathop{\mel**{r'f_{0}}{\qty( - \frac{1}{2} [X,Y] )}{rf_{0}-\sqrt{\frac{\omega_{\UV}}{\omega_{\Lambda}}}g}}}\notag\\
&\begin{aligned}
    =\frac{1}{2}\sideset{_{\UV}}{}{\mathop{\Bra{r'f_{0}\vphantom{\frac{\omega_{\UV,1}}{\omega_{\Lambda,1}}}}}}\Bigg[&-\frac{\lambda}{2}\int_{p,\ell,k_2 k_3}\sqrt{\frac{\omega_{\UV,p}}{\omega_{\Lambda,p}}}g(-p)g(-\ell)C_{2}(p, k_2; k_3; \ell)a^{+}_{\UV,-k_2}a^{+}_{\UV,-k_3}\\
    &- \frac{\lambda}{2} \int_{p,k}\sqrt{\frac{\omega_{\UV,p}}{\omega_{\Lambda,p}}}g(-p)g(p) C_{3}(k; -p)\Bigg]\Ket{rf_{0}-\sqrt{\frac{\omega_{\UV}}{\omega_{\Lambda}}}g}_{\UV}
\end{aligned}\\
&\begin{aligned}
    &= - \frac{\lambda}{4} \int_{p,\ell,k_2,k_3} \sqrt{\frac{\omega_{\UV,p}}{\omega_{\Lambda,p}}} g(-p) g(-\ell) C_{2}(p, k_2 ; k_3; \ell) \\
    &\quad\quad\quad \times \sideset{_{\UV}}{}{\mathop{\Bra{r'f_{0}\vphantom{\frac{\omega_{\UV,1}}{\omega_{\Lambda,1}}}}}}\left[a_{\UV,-k_2} a_{\UV,-k_3} + a_{\UV,k_2}^{\dagger} a_{\UV,-k_3} - a_{\UV,k_2}^\dagger a_{\UV,k_3}^\dagger  - a_{\UV,k_3}^\dagger a_{\UV,-k_2} \right. \\
    &\quad\quad\quad \left. - \tilde{\delta}(-k_2 - k_3) \right] \Ket{rf_{0} - \sqrt{\frac{\omega_{\UV}}{\omega_{\Lambda}}} g }_{\UV} \\
    &\quad\quad - \frac{\lambda}{4} \int_{p,k} \sqrt{\frac{\omega_{\UV,p}}{\omega_{\Lambda,p}}} g(-p) g(p) C_{3}(k;-p) \sideset{_{\UV}}{_{\UV}}{\mathop{\Braket{r'f_{0}|rf_{0}-\sqrt{\frac{\omega_{\UV}}{\omega_{\Lambda}}}g}}}
\end{aligned}\\
&\begin{aligned}
    &= -\frac{\lambda}{4} \int_{p,l,k_2,k_3} \sqrt{\frac{\omega_{\UV,p}}{\omega_{\Lambda,p}}} g(-p) g(-\ell) C_{2}(p,k_2;k_3;\ell) \\
    &\quad \times \left\{ \qty( r^2 - r'^2 ) f_{0}(-k_2) f_{0}(-k_3) - (r + r') \sqrt{\frac{\omega_{\UV,3}}{\omega_{\Lambda,3}}} f_{0}(-k_2) g(-k_3) \right. \\
    &\quad \quad \quad - (r - r') \sqrt{\frac{\omega_{\UV,2}}{\omega_{\Lambda,2}}} f_{0}(-k_3) g(-k_2) + \sqrt{\frac{\omega_{\UV,2} \omega_{\UV,3}}{\omega_{\Lambda,2} \omega_{\Lambda,3}}} g(-k_2) g(-k_3) \Bigg\}\\
    &\qquad\times\sideset{_{\UV}}{_{\UV}}{\mathop{\Braket{r'f_{0}|rf_{0}-\sqrt{\frac{\omega_{\UV}}{\omega_{\Lambda}}}g}}}. \\
\end{aligned}
\end{align}
The Y term is 
\begin{align}
&\sideset{_{\UV}}{_{\UV}}{\mathop{\mel**{r'f_{0}}{Y}{rf_{0}-\sqrt{\frac{\omega_{\UV}}{\omega_{\Lambda}}}g}}}\notag\\
&\begin{aligned}
    =\sideset{_{\UV}}{}{\mathop{\Bra{r'f_{0}\vphantom{\frac{\omega_{\UV,1}}{\omega_{\Lambda,1}}}}}}\int_p g(-p)\Bigg[\bigg\{
    &- \frac{\lambda}{4!} \int _{k_1 k_2 k_3} C_{1}(k_1, k_2, k_3; p) a^{-}_{\UV,-k_1}a^{-}_{\UV,-k_2}a^{-}_{\UV,-k_3}\\
    &- \frac{\lambda}{8} \int_{k_1 k_2 k_3} C_{2}(k_1, k_2; k_3; p)a^{+}_{\UV,-k_1}a^{+}_{\UV,-k_2}a^{-}_{\UV,-k_3}\\
    &- \frac{\lambda}{4} \int_{k} C_{3}(k; p)a^{+}_{\UV,p}
    - C_{4}(p)a^{-}_{\UV,p}\bigg\}\Bigg]\Ket{rf_{0}-\sqrt{\frac{\omega_{\UV}}{\omega_{\Lambda}}}g}_{\UV}.
\end{aligned}
\end{align}
We evaluate each term separately.
\paragraph{$C_{1}$ term}\mbox{}\\
    First, we expand $a^{\pm}_{\UV}$ by creation and annihilation operators, and then bring the creation operators to the left and the annihilation operators to the right. Then, we obtain
        \begin{align}
            a^{-}_{\UV,-k_1} a^{-}_{\UV,-k_2} a^{-}_{\UV,-k_3}
            &= \left( a_{\UV,k_1} - a^{\dag}_{\UV,k_1} \right) \left( a_{\UV,k_2} - a^{\dag}_{\UV,k_2} \right) \left( a_{\UV,k_3} - a^{\dag}_{\UV,k_3} \right) \nonumber \\
            &\begin{aligned}
                &\rightarrow a_{\UV,-k_1} a_{\UV,-k_2} a_{\UV,-k_3} - 3 a^{\dag}_{\UV,k_1} a_{\UV,-k_2} a_{\UV,-k_3} \\
                &\quad\quad + 3 a^{\dag}_{\UV,k_1} a^{\dag}_{\UV,k_2} a_{\UV,-k_3} - a^{\dag}_{\UV,k_1} a^{\dag}_{\UV,k_2} a^{\dag}_{\UV,k_3} \\
                &\quad\quad - 3\tilde{\delta}(-k_{1} - k_{3}) a_{\UV,-k_2} + 3\tilde{\delta}(-k_{1} - k_{3}) a^{\dag}_{\UV,k_2} \ .
            \end{aligned}
        \end{align}
        Here, the right arrow means that the terms that become identical due to the replacement of the integral variables are lumped together. Then,
        \begin{align}
            &- \frac{\lambda}{4!}\int_{p,k_1 , k_2, k_3} g(-p) C_{1}(k_1, k_2, k_3; p) \sideset{_{\UV}}{_{\UV}}{\mathop{\mel**{r'f_{0}}{a^{-}_{\UV,-k_1}a^{-}_{\UV,-k_2}a^{-}_{\UV,-k_3}}{rf_{0}-\sqrt{\frac{\omega_{\UV}}{\omega_{\Lambda}}}g}}}\notag\\
            &\begin{aligned}
                &= - \frac{\lambda}{4!} \int_{p,k_1,k_2,k_3} g(-p) C_{1} (k_1, k_2, k_3 ; p) \\
                &\quad \times \left\{ (r - r')^3 f_{0}(-k_1) f_{0}(-k_2) f_{0}(-k_3) - 3 (r - r')^2 \sqrt{\frac{\omega_{\UV,3}}{\omega_{\Lambda,3}}} f_{0}(-k_1) f_{0}(-k_2) g(-k_3) \right. \\
                &\qquad\qquad\left. + 3 (r - r') \sqrt{\frac{\omega_{\UV,2} \omega_{\UV,3}}{\omega_{\Lambda,2} \omega_{\Lambda,3}}} f_{0}(-k_1) g(-k_2) g(-k_3)\right.\\
                &\left.\qquad\qquad- \sqrt{\frac{\omega_{\UV,1}}{\omega_{\Lambda,1}}} g(-k_1) \sqrt{\frac{\omega_{\UV,2}}{\omega_{\Lambda,2}}} g(-k_2) \sqrt{\frac{\omega_{\UV,3}}{\omega_{\Lambda,3}}} g(-k_3) \right\}\\ 
                &\qquad\qquad\times\sideset{_{\UV}}{_{\UV}}{\mathop{\Braket{r'f_{0}|rf_{0}-\sqrt{\frac{\omega_{\UV}}{\omega_{\Lambda}}}g}}} \\
                &+ \frac{\lambda}{8} \int_{p,k_1,k_2} g(-p) C_{1} (k_1, k_2, -k_1 -k_2 ; p) \left\{ (r - r') f_{0}(-k_2) - \sqrt{\frac{\omega_{\UV,2}}{\omega_{\Lambda,2}}} g(-k_2) \right\}\\
                &\qquad\times\sideset{_{\UV}}{_{\UV}}{\mathop{\Braket{r'f_{0}|rf_{0}-\sqrt{\frac{\omega_{\UV}}{\omega_{\Lambda}}}g}}} \ .
            \end{aligned}
    \end{align}
\paragraph{$C_{2}$ term}\mbox{}\\
        By following the same process used for the $C_1$ term, we obtain
        \begin{align}
                &\begin{aligned}
                    a^{+}_{\UV,-k_1} a^{+}_{\UV,-k_2} a^{-}_{\UV,-k_3}
                    &\rightarrow a_{\UV,-k_1} a_{\UV,-k_2} a_{\UV,-k_3} - a^{\dag}_{\UV,k_3} a_{\UV,-k_1} a_{\UV,-k_2} \\
                    &\quad + 2 a^{\dag}_{\UV,k_1} a_{\UV,-k_2} a_{\UV,-k_3} - 2 a^{\dag}_{\UV,k_1} a^{\dag}_{\UV,k_3} a_{\UV,-k_2} \\
                    &\quad + a^{\dag}_{\UV,k_1} a^{\dag}_{\UV,k_2} a_{\UV,-k_3} - a^{\dag}_{\UV,k_1} a^{\dag}_{\UV,k_2} a^{\dag}_{\UV,k_3} \\
                    &\quad - 2 \tilde{\delta}(-k_{1} - k_{3}) a_{\UV,-k_2} + \tilde{\delta}(-k_{1} - k_{2}) a_{\UV,-k_3} \\
                    &\quad - \tilde{\delta}(-k_{1} - k_{2}) a^{\dag}_{\UV,k_3} - 2 \tilde{\delta}(-k_{2} - k_{3}) a^{\dag}_{\UV,k_1} \ .
                \end{aligned}
        \end{align}
        Again, the right arrow means that the terms that become identical due to the replacement of the integral variables are lumped together.
        Then, 
        \begin{align}
            &- \frac{\lambda}{8} \int_{p, k_1, k_2, k_3} g(-p)C_{2}(k_1, k_2; k_3; p)\sideset{_{\UV}}{}{\mathop{\Bra{r'f_{0}\vphantom{\frac{\omega_{\UV,1}}{\omega_{\Lambda,1}}}}}}a^{+}_{\UV,-k_1}a^{+}_{\UV,-k_2}a^{-}_{\UV,-k_3}\Ket{rf_{0}-\sqrt{\frac{\omega_{\UV}}{\omega_{\Lambda}}}g}_{\UV}\notag\\
                &\begin{aligned}
                &= - \frac{\lambda}{8} \int_{p,k_1,k_2,k_3} g(-p) C_2(k_1, k_2; k_3; p) \\
                &\quad \times \left\{\vphantom{\sqrt{\frac{\omega_{\UV,3}}{\omega_{\Lambda,3}}}} (r - r') (r^2 + r'^2) f_{0}(-k_1) f_{0}(-k_2) f_{0}(-k_3) \right. - 2 (r^2 - r'^2) \sqrt{\frac{\omega_{\UV,1}}{\omega_{\Lambda,1}}} g(-k_1) f_{0}(-k_2) f_{0}(-k_3) \\
                &\quad\quad - (r + r')^2 \sqrt{\frac{\omega_{\UV,3}}{\omega_{\Lambda,3}}} f_{0}(-k_1) f_{0}(-k_2) g(-k_3)  + 2 (r + r') \sqrt{\frac{\omega_{\UV,2} \omega_{\UV,3}}{\omega_{\Lambda,2} \omega_{\Lambda,3}}} f_{0}(-k_1)g(-k_2) g(-k_3) \\
                &\quad\quad + (r - r') \sqrt{\frac{\omega_{\UV,1} \omega_{\UV,2}}{\omega_{\Lambda,1} \omega_{\Lambda,2}}} g(-k_1) g(-k_2) f_{0}(-k_3) \left. - \sqrt{\frac{\omega_{\UV,1} \omega_{\UV,2} \omega_{\UV,3}}{\omega_{\Lambda,1} \omega_{\Lambda,2} \omega_{\Lambda,3}}} g(-k_1) g(-k_2) g(-k_3) \right\}   \\
                &\quad + \frac{\lambda}{4} \int_{p,k_1,k_2,k_3} g(-p) C_2(k_1, k_2; -k_1; p) \\
                &\quad\quad \times \left\{ (r + r') f_{0}(-k_2) - \sqrt{\frac{\omega_{\UV,2}}{\omega_{\Lambda,2}}} g(-k_2) \right\} \sideset{_{\UV}}{_{\UV}}{\mathop{\Braket{r'f_{0}|rf_{0} - \sqrt{\frac{\omega_{\UV}}{\omega_{\Lambda}}}g}}} \\
                &\quad - \frac{\lambda}{8} \int_{p,k_1,k_2,k_3} g(-p) C_2(k_1, -k_1; k_3; p) \\
                &\quad\quad \times \left\{ (r - r') f_{0}(-k_3) - \sqrt{\frac{\omega_{\UV,3}}{\omega_{\Lambda,3}}} g(-k_3) \right\} \sideset{_{\UV}}{_{\UV}}{\mathop{\Braket{r'f_{0}|rf_{0} - \sqrt{\frac{\omega_{\UV}}{\omega_{\Lambda}}}g}}} \ .
            \end{aligned}
        \end{align}
\paragraph{$C_{3}$ and $C_{4}$ term}
        \begin{align}
            &\begin{aligned}
            &- \frac{\lambda}{4} \int_{p, k} C_3(k; p)\sideset{_{\UV}}{}{\mathop{\Bra{r'f_{0}\vphantom{\frac{\omega_{\UV,1}}{\omega_{\Lambda,1}}}}}}
            a^{+}_{\UV,p}\Ket{rf_{0}-\sqrt{\frac{\omega_{\UV}}{\omega_{\Lambda}}}g}_{\UV}\\
            &\hphantom{- \frac{\lambda}{4} \int_{p, k} C_3(k; p)\sideset{_{\UV}}{}{\mathop{\Bra{r'f_{0}\vphantom{\frac{\omega_{\UV,1}}{\omega_{\Lambda,1}}}}}}
            a^{+}_{\UV,p}}
            -\int_{p}g(-p) C_4(p)\sideset{_{\UV}}{}{\mathop{\Bra{r'f_{0}\vphantom{\frac{\omega_{\UV,1}}{\omega_{\Lambda,1}}}}}}
            a^{-}_{\UV,p}\Ket{rf_{0}-\sqrt{\frac{\omega_{\UV}}{\omega_{\Lambda}}}g}_{\UV}
            \end{aligned}
            \notag\\
            &\begin{aligned}
                &= - \frac{\lambda}{4} \int_{p,k} g(-p) C_3(k; p) \left\{ \qty(r+r') f_{0}(p) - \sqrt{\frac{\omega_{\UV,p}}{\omega_{\Lambda,p}}} g(p) \right\} \sideset{_{\UV}}{_{\UV}}{\mathop{\Braket{r'f_{0}|rf_{0} - \sqrt{\frac{\omega_{\UV}}{\omega_{\Lambda}}}g}}} \\
                &\quad - \int_p g(-p) C_4(p) \left\{ \qty(r-r') f_{0}(p) - \sqrt{\frac{\omega_{\UV,p}}{\omega_{\Lambda,p}}} g(p) \right\}\sideset{_{\UV}}{_{\UV}}{\mathop{\Braket{r'f_{0}|rf_{0} - \sqrt{\frac{\omega_{\UV}}{\omega_{\Lambda}}}g}}} \ .
            \end{aligned}
        \end{align}
In this way, we have finished calculating the Y term. Lastly, combining all the above calculations, we obtain (\ref{error exp value 2}).

\section*{Appendix E: Scaling of $\delta m^2_{\Lambda}$}
\renewcommand{\theequation}{E.\arabic{equation}}
\setcounter{equation}{0}
In this section, in order to calculate the $\Lambda/m$ dependence of $\delta m^2_{\Lambda}$, we solve \eqref{scaling eq for delta m^2}.
First, we set $\delta m^2_{\Lambda}$ to
\begin{align}
    \label{solution of scaling eq for delta m^2}
    \delta m^2_{\Lambda} = \qty(\frac{\Lambda_{\UV}}{\Lambda})^2C_{\Lambda} \ .
\end{align}
By substituting this into \eqref{scaling eq for delta m^2}, we obtain
\begin{align}
    -\Lambda\frac{dC_{\Lambda}}{d\Lambda} = -\frac{\lambda_{\Lambda}}{2}\qty(\frac{\Lambda}{\Lambda_{\UV}})^2\int_p\qty(\frac{d}{2\omega_{\Lambda,p}}-\frac{p^2}{2\omega^{3}_{\Lambda,p}})K_p \ .
\end{align}
Then, we need to calculate
\begin{align}
    C_{\Lambda} = \frac{1}{\Lambda_{\UV}^2}\int_p\frac{K_p}{2}\int^{\Lambda}\Lambda'd\Lambda'\lambda_{\Lambda'}\qty(\frac{d}{2\omega_{\Lambda',p}}-\frac{p^2}{2\omega^{3}_{\Lambda',p}}) \ .
\end{align}
Here, $\lambda$ does not vary with $\Lambda$ to the first order of $\alpha$, so that $\lambda$ can be factored out of the integral.
The result is
\begin{align}
\begin{aligned}
    \label{C_Lambda}
    C_{\Lambda} = 
    \frac{\lambda}{\Lambda_{\UV}^2}\int_p\frac{K_p}{2}&\left[-\frac{m^2}{2p^2}\frac{1}{\omega_{\Lambda,p}}
    +\frac{(d-1)m^2}{8p^2}\qty(\frac{1}{\omega_{\Lambda,p}+p}-\frac{1}{\omega_{\Lambda,p}-p})\right.\\
    &\qquad\qquad\left.\vphantom{\frac{\abs{\omega_{\Lambda,p}-p}}{\abs{\omega_{\Lambda,p}+p}}}
    +\frac{(d-3)m^2}{8p^3}\ln\qty(\frac{\omega_{\Lambda,p}-p}{\omega_{\Lambda,p}+p})\right] + (\text{integration constant}) \ .
\end{aligned}
\end{align}
Now, let us see 
the behavior of $\delta m^2_{\Lambda}$ at small $\Lambda/m$. 
The first and second terms in \eqref{C_Lambda} obviously behaves as $\Lambda/m$. The third term also behaves as
\begin{align}
    \ln\qty(\frac{\omega_{\Lambda,p}-p}{\omega_{\Lambda,p}+p}) \sim \frac{\Lambda}{m} \ .
\end{align}
We therefore see from \eqref{solution of scaling eq for delta m^2} and \eqref{C_Lambda} that $\delta m^2_{\Lambda}$ behaves as $(\Lambda/m)^{-2}$ at small $\Lambda/m$, which comes from the integration constant in \eqref{C_Lambda}.

\bibliographystyle{ptephy_arxiv}
  \bibliography{ref}

\end{document}